\documentclass[printer]{aa}
\usepackage{graphicx}
\usepackage{txfonts}

\usepackage[utf8]{inputenc}
\bibpunct{(}{)}{;}{a}{}{,}
\usepackage{graphicx,natbib,url,twoopt,subcaption}
\usepackage[T1]{fontenc}
\usepackage{rotating,booktabs}
\usepackage{float}

\usepackage{adjustbox}
\usepackage{xcolor}
\usepackage{amsmath} 
\usepackage{gensymb}
\usepackage{placeins}
\usepackage{lineno}
\usepackage{soul}

\begin{document}    

   \title{Spectral analysis of two directly imaged benchmark L dwarf companions at the stellar-substellar boundary \thanks{Based on observations collected with SPHERE mounted on the VLT at Paranal Observatory (ESO, Chile) under programs 0103.C-0199(A) (PI: Rickman), and 105.20SZ.001 (PI: Rickman).}}

   \author{W. Ceva\inst{\ref{obsge}}
          \and
          E. C. Matthews\inst{\ref{mpia}, \ref{obsge}}
          \and
          E. L. Rickman\inst{\ref{stsci}, \ref{obsge}}
          \and
          D. S\'egransan\inst{\ref{obsge}}
          \and
          A. Vigan\inst{\ref{marseille}}
          \and
           B. P. Bowler\inst{\ref{santabarbara}}
          \and
          T. Forveille\inst{\ref{grenoble}}
          \and
          K. Franson\inst{\ref{texas}}
          \and
          J. Hagelberg\inst{\ref{obsge}}
          \and
          S. Udry\inst{\ref{obsge}} 
          }

   \institute{D\'{e}partment d'astronomie de l’Universit\'{e} de Gen\`{e}ve, Chemin Pegasi 51, 1290 Versoix, Switzerland \\ \email{william.ceva@unige.ch}\label{obsge}
   \and
   Max-Planck-Institut f\"{u}r Astronomie, Königstuhl 17, D-69117 Heidelberg, Germany\label{mpia}
   \and
   European Space Agency (ESA), ESA Office, Space Telescope Science Institute, 3700 San Martin Drive, 
    Baltimore 21218, MD, USA\label{stsci}
   \and
   Aix Marseille Universit\'{e}, CNRS, CNES, LAM, Marseille, France \label{marseille}
   \and
   Department of Physics, University of California, Santa Barbara, Santa Barbara, CA 93106, USA
   \label{santabarbara}
   \and
   Universit\'{e} Grenoble Alpes, CNRS, IPAG, 38000 Grenoble, France\label{grenoble}
   \and
   Department of Astronomy, The University of Texas at Austin, Austin, TX 78712, USA\label{texas}}

   \date{Received --; accepted --}
   \authorrunning{Ceva et al.}
   \titlerunning{Spectral analysis of two directly imaged benchmark L dwarf companions at the stellar-substellar boundary}

  \abstract
   {We used multiple epochs of high-contrast imaging spectrophotometric observations to determine the atmospheric characteristics and thermal evolution of two previously detected benchmark L dwarf companions, HD~112863~B and HD~206505~B.  We analyzed IRDIS and IFS data from VLT/SPHERE of each companion, both of which have dynamical masses near the stellar-substellar boundary.  We compared each companion with empirical spectral standards, as well as constrained their physical properties through atmospheric model fits.  From these analyses, we estimate that HD~112863~B is spectral type L$3\pm1$ and that HD~206505~B is spectral type L$2\pm1$.  Using the BT-Settl atmospheric model grids, we find a bimodal solution for the atmospheric model fit of HD~112863~B, such that $T_{\rm{eff}}=1757^{+37}_{-36}$~K or $2002^{+23}_{-24}$~K and $\log{g}=4.973^{+0.057}_{-0.063}$ or $5.253^{+0.037}_{-0.033}$, while for HD~206505~B, $T_{\rm{eff}}=1754^{+13}_{-13}$~K and $\log{g}=4.919^{+0.031}_{-0.029}$.  Comparing the bolometric luminosities of both companions with evolutionary models imply that both companions are likely above the hydrogen burning limit.}

   \keywords{planets and satellites: detection -- techniques: high angular resolution -- stars: imaging, low-mass, individual -- HD~112863, HD~206505}

   \maketitle

\section{Introduction} \label{sec:intro}

Brown dwarfs (BDs) and very low mass stars (VLMSs) can collectively be referred to as ultra-cool dwarfs (UCDs).  These objects comprise the M, L, T, and Y spectral types, and roughly correspond to effective temperatures $T_{\rm{eff}}\lesssim3000$~K.  The latest possible occurrence of the stellar-substellar boundary, which differentiates VLMSs and BDs, is in the L spectral type (e.g. \citealp{Burrows2001}, \citealp{Dieterich2014}, \citealp{Dupuy2017}); VLMSs can be either type M or L, while all objects of T type or lower \citep[$T_{\rm{eff}}\lesssim 1350$~K,][]{Burgasser2002, Cushing2006, Cushing2008, Dupuy2017} must be substellar, since an object with $T_{\rm{eff}}\lesssim 1600$~K at hydrostatic equilibrium cannot fuse hydrogen (\citealp{Chabrier2000}, \citealp{Chabrier_etal2000}, \citealp{Baraffe2003}, \citealp{Dieterich2014}, \citealp{Zhang2017}, \citealp{Chabrier2023}).  The L spectral type therefore includes objects which do not follow the typical mass-luminosity-age distinction of main sequence stars, since BDs cool and contract as they age, leading to a degeneracy between these three parameters (\citealp{Martin1997}, \citealp{Kirkpatrick1999}, \citealp{Kirkpatrick2005}).  For example, a young, low mass L-type BD can have the same luminosity as an older, higher mass L-type VLMS.  Distinguishing between these two scenarios requires independent measurements of these fundamental parameters.

The stellar-substellar boundary, also known as the hydrogen-burning limit (HBL), is the limit below which the bolometric luminosity $L$ of an UCD will never be equivalent to power generated by hydrogen fusion in its core.  In other words, it is the limit below which UCDs will not reach a stable $L$, but will instead continue to cool throughout their lifetimes (\citealp{Burrows2001}, \citealp{Chabrier2023}).  The HBL is determined by the hydrogen-burning minimum mass (HBMM), which is the lowest mass $M$ at which an object in hydrostatic equilibrium can still achieve a core temperature $T_{\rm{c}}$ where the kinetic energy of protons is high enough to overcome the Coulomb barrier and fully initiate the proton-proton chain \citep{deBoer2008}.  The HBMM is metallicity and helium mass fraction dependent, since a higher metallicity and/or higher helium mass fraction in the core of an UCD increases the mean molecular weight and therefore $T_{\rm{c}}$, such that the HBMM is inversely proportional to metallicity and helium mass fraction \citep{Chabrier1997, Chabrier2000, Burrows2001, Burrows2011, Zhang2017, Marley2021}.  The HBMM is also dependent on clouds, specifically the opacity of the grains that comprise clouds in UCD atmospheres; higher cloud opacity/thickness both increases $T_{\rm{c}}$ by raising the temperature differential throughout the interior of an UCD, and decreases the $L$ needed to reach the main sequence by impeding photon escape from the UCD surface (\citealp{Chabrier_etal2000}, \citealp{Burrows2001}, \citealp{Saumon2008}).  Thus, determining whether an UCD lies above or below the HBL requires constraints on $M$, as well as metallicity, helium mass fraction, and cloud/grain opacity.  Model predictions of the HBMM vary: \cite{Chabrier1997} predicts a HBMM of $\sim 0.072$~$M_{\odot}$ to $\sim0.083$~$M_{\odot}$ depending on metallicity, \cite{Saumon2008} predicts a HBMM of  $0.070$~$M_{\odot}$ for cloudy UCDs and $0.075$~$M_{\odot}$ without clouds, \cite{Fernandes2019} finds a range of HBMMs of  $0.074$~$M_{\odot}$ to $0.080$~$M_{\odot}$ depending on metallicity, \cite{Marley2021} predicts a HBMM of $0.074$~$M_{\odot}$ at solar metallicity, \cite{Chabrier2023} predicts a HBMM of $0.075$~$M_{\odot}$ at solar abundances, and \cite{Morley2024} finds a range of HBMMs of $0.063$~$M_{\odot}$ to $0.078$~$M_{\odot}$ depending on metallicity and cloud presence.

Testing such predictions made by evolutionary models can be accomplished by comparing, for instance, a model-independent measurement of an UCD's $M$, with a model-derived $M$ value based on observations of an UCD's luminosity.  A model-independent $M$ of an UCD can be acquired via a measurement of an UCD's dynamical mass.  Such UCDs, which are companions to stars and/or another UCD, are referred to as ``benchmark'' UCDs (e.g. \citealp{Bouy2003}, \citealp{Calamari2024}).  Many detections of such objects can be found in the literature, with each new detection providing an additional opportunity to test the underlying physics of evolutionary and atmospheric models of UCDs \citep[e.g.][]{Lane2001, Liu2008, Crepp2014, Crepp2016, Cheetham2018, Rickman2020, Brandt2020, Currie2020, Venner2021, Rickman2022, Liu2025}.

Two recently discovered benchmark UCDs are HD~112863~B and HD~206505~B, which were detected in high-contrast imaging (HCI) observations by \cite{Rickman2024}.  That work found that the dynamical mass of HD~112863~B is $M=77.1^{+2.9}_{-2.8}$~$M_{\rm{Jup}}$ and that the dynamical mass of HD~206505~B is $M=79.8\pm1.8$~$M_{\rm{Jup}}$.  These both lie within the range of $M\sim0.063-0.083$~$M_{\odot}$ ($M\sim66-87$~$M_{\rm{Jup}}$), where evolutionary models predict the HBMM (\citealp{Chabrier1997}, \citealp{Morley2024}).  Furthermore, the photometric analysis from \cite{Rickman2024} show that both HD~112863~B and HD~206505~B are L dwarfs.  

This paper presents the spectral analysis of HD~112863~B and HD~206505~B, complementing the photometry and dynamical masses derived in \cite{Rickman2024}.   The spectra of both L dwarfs were collected simultaneously with the photometry data presented in \cite{Rickman2024}.  Spectral analysis of UCDs is important since it provides constraints on UCD atmospheric properties, such as $T_{\rm{eff}}$, surface gravity $\log{g}$, metallicity, cloud properties, etc. (e.g. \citealp{Chabrier_etal2000}, \citealp{Hiranaka2016}, \citealp{Charnay2018}, \citealp{Mukherjee2022}, \citealp{Godoy2024}, \citealp{Lewis2024}, \citealp{Leggett2025}).  Consequently, spectral analysis is necessary for determining whether HD~112863~B and HD~206505~B are above or below the HBL, and therefore, whether they are likely to be VLMSs or BDs.  Such analyses are useful, since UCDs with dynamical masses near the HBMM can be used to empirically constrain the HBL itself (e.g. \citealp{Dupuy2017}).  

This paper is outlined as follows: in Sect.~\ref{sec:observs} we review the HCI observations of HD~112863 and HD~206505, including the issues that occurred during the observations which complicated our data reduction steps.  In Sect.~\ref{sec:datareduc} we describe our data reduction approach, with an emphasis on non-standard steps taken in Sect.~\ref{subsubsec:weather} and Sect.~\ref{subsubsec:coron} in order to alleviate additional complications arising due to variable weather conditions and one companion being behind the coronagraph.  We note that some of the data reduction steps described in Sect.~\ref{subsubsec:weather} and Sect.~\ref{subsubsec:coron} were not included in the analysis of \cite{Rickman2024}.  As a result of this additional analysis, we provide updated relative astrometry and photometry of HD~112863~B and HD~206505~B in Appendix~\ref{appendix_relastrom}, an updated color-magnitude diagram (CMD) with both companions in Appendix~\ref{appendix_CMD}, and new orbit fits for both systems, including revised companion dynamical mass estimates, in Appendix~\ref{appendix_orbitfits}.  We analyze the extracted spectra of both UCDs, using comparisons with empirical spectra and fits to atmospheric models, in Sect.~\ref{sec:allspectrafits}, and compare the results with evolutionary models in Sect.~\ref{sec:evolmodel}.  Finally, a summary of our results is found in Sect.~\ref{sec:conclus}.  The host star properties adopted in this paper are taken from \cite{Rickman2024}.

\section{Observations} \label{sec:observs}

We first identified HD~112863 and HD~206505 as suitable HCI targets based on their long-term radial velocity (RV) trends observed as part of the CORALIE survey (\citealp{Queloz2000}, \citealp{Udry2000}).  We observed both targets with VLT/SPHERE \citep{2019A&A...631A.155B}, and presented the detections and photometric characterizations of the companions in \cite{Rickman2024}.  Here we present the spectroscopic observations that were collected simultaneously with those measurements. The SPHERE observations were conducted using SPHERE IRDIFS and IRDIFS-EXT modes, both of which use a dichroic mirror to obtain simultaneous photometric and spectroscopic data via the InfraRed Dual-Band Imager and Spectrograph \citep[IRDIS,][]{2008SPIE.7014E..3LD} and the Integral Field Spectrograph \citep[IFS,][]{2008SPIE.7014E..3EC}, respectively.

Both targets were observed in two separate epochs using the Angular Differential Imaging (ADI) observing technique \citep{2006ApJ...641..556M}.  The first epoch of observations for each target were obtained using IRDIFS mode, which provides photometry in $H2$ ($\lambda=1.5888$~$\mu$m) and $H3$ ($\lambda=1.6671$~$\mu$m) bands, along with low resolution (~$\sim50$) spectroscopy across the $YJ$ ($\lambda = 0.95 - 1.35$~$\mu$m) range.  The second epoch of observations for each target were obtained using IRDIFS-EXT mode, which provides $K1$ ($\lambda=2.1025$~$\mu$m) and $K2$ ($\lambda=2.255$~$\mu$m) photometry and $YJH$ ($0.95 - 1.65$~$\mu$m) low resolution (~$\sim30$) spectroscopy.  These follow-up IRDIFS-EXT mode observations therefore provided spectroscopy over a wider range of wavelengths, and photometric constraints on the companions' spectra farther in the infrared via the $K1$ and $K2$ bands \citep{Zurlo2014}.  Both the IRDIFS and IRDIFS-EXT observations for each target were taken using SPHERE's N\_ALC\_YJH\_S coronagraph (inner working angle IWA~$\sim0.15$~arcsec\footnote{From the VLT SPHERE User Manual, 18th release}).  
Each epoch included observations of the sky background (``sky'' frames), followed by observations of the target star's point spread function (PSF) dithered out from behind the coronagraph (``flux'' frames), observations of the target star behind the coronagraph with a pair of orthogonal $\sin$ waves imposed on the deformable mirror (``star center'' frames), and standard coronagraphic science observations.  The sky frames provide an estimate of the sky background to be subtracted from each frame, the dark current, and the positions of the bad pixels on the detector.  The flux frames provide an image of the detector PSF for use as the forward model of a companion during post-processing.  The star center frames, taken before and after the science sequence, are used to determine the position of the target star behind the coronagraph during science observations; the two orthogonal $\sin$ waves on the deformable mirror diffract the target star light such that four PSFs appear in the images, centered around the target star's position.  This is necessary in order to extract accurate relative astrometry measurements during post-processing.  

HD~112863 was observed in IRDIFS mode as part of program 105.20SZ.001 (PI: Rickman) on 2021-04-07.  This set of observations was affected by extensive clouds during the science sequence, per the Paranal Differential Image Motion Monitor (DIMM)\footnote{Paranal Astronomical Site Monitoring (ASM) data for 2021-04-07: \url{https://www.eso.org/asm/ui/publicLog?name=Paranal&startDate=2021-04-07T21:00:00.000Z&hoursInterval=15}}.  The total integration time of the science sequence in this observational epoch was $4096$~seconds with both IRDIS $H23$ and IFS $YJ$, which were both comprised of $64$~science frames, each with a detector integration time (DIT) of $64$~seconds.  Follow-up observations of HD~112863 were conducted in IRDIFS-EXT mode, also as part of program 105.20SZ.001, on 2022-01-30.  This epoch of observations was also affected by weather, with the seeing reaching $2$~arcsec during the science sequence, per the DIMM\footnote{Paranal ASM data for 2022-01-30: \url{https://www.eso.org/asm/ui/publicLog?name=Paranal&startDate=2022-01-30T22:00:00.000Z&hoursInterval=15}}.  This caused the adaptive optics (AO) loop to break at the end of the initial science sequence; a second science sequence, with its own associated star center and flux frames, was obtained in order to properly complete the observations.  The total integration time of the science sequences in this observational epoch was $4864$~seconds with both IRDIS $K12$ and IFS $YJH$, which were both comprised of $76$~science frames with DITs of $64$~seconds.

HD~206505 was observed in IRDIFS mode on 2019-08-06 as part of program 103.C-0199(A) (PI: Rickman).  This epoch of observations included less than ideal seeing, exceeding $1$~arcsec at times, according to the DIMM\footnote{Paranal ASM data for 2019-08-06: \url{https://www.eso.org/asm/ui/publicLog?name=Paranal&startDate=2019-08-06T21:00:00.000Z&hoursInterval=15}}.  The total integration time of the science sequence was $8192$~seconds with both IRDIS $H23$ and IFS $YJ$, each comprised of $128$~science frames with DITs of $64$~seconds.  Follow-up observations of HD~206505 were then obtained in IRDIFS-EXT mode on 2021-07-01 as part of program 105.20SZ.001.  Per the DIMM, this night of observations was stable, with the exception of a brief period of cloud cover lasting $<30$~minutes\footnote{Paranal ASM data for 2021-07-01: \url{https://www.eso.org/asm/ui/publicLog?name=Paranal&startDate=2021-07-01T21:00:00.000Z&hoursInterval=15}}.  The total integration time of the science sequence was $6144$~seconds with both IRDIS $K12$ and IFS $YJH$, which were comprised of $384$~science frames of $64$~second DITs. 

For each set of observations, we used the IRDIS lamp flats, as well as the IFS lamp flats (including white light, multiple narrowband, and integral field unit flats), dark frames, spectra registration images, and wavelength calibration images that were obtained closest to the time of our observations, for performing calibrations. 

\section{Data reduction} \label{sec:datareduc}

Here we describe the data reduction for both IRDIS and IFS.  We first discuss the pre-processing steps (i.e., cleaning, calibrating, and aligning the data) and then discuss our procedure to remove starlight from the images and extract spectroscopic measurements of the companions.  We also detail the additional steps we implemented to address complications unique to these datasets (i.e. unfavorable weather conditions and companion flux attenuation caused by the coronagraph).

\subsection{Pre-processing} \label{subsec:preproc}

As described in \cite{Rickman2024}, the IRDIS data was reduced using the Geneva Reduction and Analysis Pipeline for High-contrast Imaging of planetary Companions \citep[GRAPHIC,][]{2016MNRAS.455.2178H}.  GRAPHIC performs flat field correction, sky subtraction, and bad pixel correction on all frames, and then centers the flux and science frames via Fourier transforms.  GRAPHIC also performs basic frame selection on the flux cubes via $5\sigma$ clipping of frames based on PSF centering, measured flux, and PSF width.  The remaining flux frames are then corrected for any difference between the neutral density filters used for science and flux observations\footnote{The corrections for the ND filter transmission make use of the ND filter curves available at: \url{https://www.eso.org/sci/facilities/paranal/instruments/sphere/inst/filters.html}}, and scaled by exposure time to match that of the science observations.  

To reduce the IFS data, we used the \texttt{vlt-sphere} Python package \citep{Vigan2020ascl}, which employs both recipes from the ESO SPHERE pipeline \citep{2008SPIE.7019E..39P} and custom-built Python scripts.  \texttt{vlt-sphere} performs flat field correction for both the detector and wavelength-dependent sensitivity variations, spectra position determination and wavelength calibration, standard dark and sky subtraction, as well as bad pixel correction.  

One notable improvement to the data reduction procedure included in \texttt{vlt-sphere} that is not a part of the ESO SPHERE pipeline is the correction for spectral crosstalk, a phenomenon described in detail in \cite{Antichi2009}.  Another improvement provided by \texttt{vlt-sphere} is using the star center frames to correct the wavelength estimation produced by the ESO SPHERE pipeline, which may determine the wavelengths of the spectra with an inaccuracy of $\sim20$~nm or higher.  Both of these improvements included in \texttt{vlt-sphere} are described in detail in \cite{Vigan2015}.  

As with GRAPHIC reductions of the IRDIS data, all IFS image shifts and centering performed by \texttt{vlt-sphere} utilize Fourier transforms.  Furthermore, \texttt{vlt-sphere} also accounts for the difference in exposure time and neutral density filters between the flux and science frames.  Finally, we applied the same GRAPHIC frame selection procedure that was applied to the IRDIS flux cubes to the IFS flux cubes, after pre-processing with \texttt{vlt-sphere} was complete.

We manually removed certain frames in the datasets before conducting pre-processing with GRAPHIC and \texttt{vlt-sphere}.  This included the first sky frame in each IRDIS $H23$ dataset, since each of these frames showed signs of of persistence caused by the preceding flux frame observations.  We also excluded the second star center cube from the 2022-01-30 IRDIS $K12$ and IFS $YJH$ datasets of HD~112863, since the target star was not behind the coronagraph in any of these frames due to the AO loop break (see Sect.~\ref{sec:observs}).  Additionally, after pre-processing with GRAPHIC and \texttt{vlt-sphere}, we removed the two frames from the science master cube in the 2022-01-30 HD~112863 IRDIS $K12$ and IFS $YJH$ datasets, where the target star was not behind the coronagraph due to the AO loop break (see Sect.~\ref{sec:observs}).  

\subsection{Photometry and spectra extraction} \label{subsec:spectraextract}

We ran the Temporal Reference Analysis of Planets \citep[TRAP,][]{Samland2021} post-processing algorithm in order to extract the contrast spectra of HD~112863~B and HD~206505~B from the pre-processed IRDIS and IFS data.  TRAP differs from many other post-processing methods in that it models systematics, particularly speckle noise, from a temporal perspective instead of a spatial perspective.  Specifically, it models pixels potentially containing a companion as a combination of 1.) a ``light curve'' of a companion ``transiting'' through the pixels due to the ADI sequence and 2.) the light curves of pixels not affected by a companion's flux, to provide an estimate of the systematic/speckle noise.  More details can be found in \cite{Samland2021}.  With TRAP, we detected HD~112863~B with a maximum SNR of ~$72$, in IFS $YJH$, while we detected HD~206505~B with a maximum SNR of ~$778$, in IFS $YJ$.

To convert the contrast spectrum to a flux spectrum also requires a stellar spectrum at the wavelengths and resolution of the IFS data. We therefore created a model stellar spectrum with uncertainties based on the physical properties of each star as derived in \cite{Rickman2024}. We used the posteriors of the stellar parameters analysis from that work, which provide values for $T_{\rm{eff}}$, $\log{g}$, host star radius, parallax $\varpi$, and metallicity $\rm{[Fe/H]}$, as well as the correlation between those parameters. We randomly drew 10000~samples from the posterior, and for each sample generated a synthetic spectrum. To do this, we used the BT-NextGen models \citep{Allard2012}; we first resampled these spectra to the needed resolution, then interpolated over $T_{\rm{eff}}$, $\log{g}$, and $\rm{[Fe/H]}$, and finally scaled the overall flux based on the radius and distance of the target. Our final stellar model with uncertainties is the median and 1$\sigma$ confidence interval at each wavelength point of the 10000~spectra generated from the stellar parameters posterior. To verify the validity of the spectrum, we repeated the same process to generate synthetic photometry in the TYCHO \citep{Hog2000}, \textit{Gaia} \citep{GaiaCollaboration2023}, 2MASS \citep{Cutri2003}, and WISE \citep{Cutri2021} bandpasses, and confirmed that these are in good agreement with the measured photometry for each star. This spectrum could then be multiplied by the contrast spectrum, to derive a flux-calibrated spectrum of the companion.

The resulting flux spectra of HD~112863~B and HD~206505~B are shown in Fig.~\ref{fig:hd112863noDTTSnoCC} and Fig.~\ref{fig:hd206505noDTTS}.  There is a clear offset in flux between the first epoch of data ($H23$/$YJ$) and the second epoch of data ($K12$/$YJH$) for both companions, with this offset being larger for HD~112863~B.  These offsets were resolved by accounting for bad weather during each epoch of observations (see Sect.~\ref{sec:observs}), and, for HD~112863~B specifically, by correcting for the flux attenuation due to the coronagraph.  

\subsubsection{Weather-driven frame selection} \label{subsubsec:weather}

According to the information provided by the Paranal DIMM, the weather deteriorated at some point during each set of observations of both companions (see Sect.~\ref{sec:observs}).  Although GRAPHIC applies frame selection on the final flux cubes to account for weather changes (see Sect.~\ref{subsec:preproc}), this does not address any differences in weather conditions between the flux frames and the science frames.  For instance, if there are clouds present during the flux observations that are not present during the science observations, then the forward model of the companion PSF would not properly represent the flux of the target star during the science sequence; the extracted contrast of the companion would be lower than the true contrast.  Conversely, if there are clouds present during the science observations but not the flux observations, the extracted contrast of the companion would be higher than the true contrast.  

In order to obtain a better understanding of the weather conditions during each epoch of observations, we analyzed the AO data recorded on the wavefront sensor (WFS) and the dedicated differential tip-tilt sensor (DTTS) inside SPHERE by the Standard Platform for Adaptive optics Real Time Applications (SPARTA) computer \citep{SuarezValles2012}.  Since this data is recorded by the SPHERE extreme AO (SAXO) system \citep{Fusco2006, Petit2012, Petit2014, Sauvage2016}, which operates at $1380$~Hz, it is taken at a cadence of $\approx 2$~seconds, much quicker than the DIMM data, which is taken at a cadence of $\approx 1.5$~minutes.  Therefore, this data provides weather information for nearly every frame in our observations.  Additionally, since this data is recorded by SAXO itself, it includes only the weather events directly impacting the instrument FOV, unlike the all-sky DIMM data.

We reduced the SPARTA files associated with each epoch of observations using the \texttt{vlt-sphere} package.  \texttt{vlt-sphere} extracts the weather data from these files, including the seeing, coherence time, strehl ratio, ground layer fraction, and wind speed as estimated by SPARTA and DIMM.  Furthermore, the DTTS records images of the target star PSF as it appears behind the coronagraph, to ensure that the target star remains centered under the coronagraph during science observations.  Therefore, \texttt{vlt-sphere} also provides the images of the PSF taken by the DTTS every $\approx 2$~seconds, as well as the measurements of the PSF flux taken by the DTTS and the WFS subapertures every $\approx 30$~seconds\footnote{From the VLT SPHERE User Manual, 18th release}.  While the WFS is sensitive to visible light, the DTTS operates in $H$~band, and thus provides a better metric for the effects of weather on our science observations.  

We therefore accounted for weather variations during our observations by applying simple frame selection to our science master cubes, based on each science frame's associated DTTS flux value(s).  In particular, we excluded science frames where the simultaneous DTTS flux value(s) were beyond $\pm1$ the median absolute deviation (MAD) from the median DTTS flux value of the entire observational epoch (see Appendix~\ref{appendix_dtts}).  This threshold was chosen since it is less affected by the DTTS flux values taken during strong cloud coverage, as opposed to a threshold of $\pm1\sigma$, which would lead to more science frames with cloud coverage being included in the final science master cube.  To determine which DTTS flux values were associated with which science frame, we assumed that the timestamp of each value is instantaneous; \texttt{vlt-sphere} only provided one time value for each DTTS flux recording, with no information on duration.  For some science frames, there were $2$~flux values recorded within the duration of the frame - in these cases, we kept the science frame only if both DTTS flux values were within $\pm1$~MAD.  Conversely, for the 2021-07-01 epoch of HD~206505 observations, some science frames had no DTTS flux values temporally associated with them, due to the higher number of science frames taken during this epoch ($384$~$K12$ and $YJH$ frames versus $246$~DTTS values, see Sect.~\ref{sec:observs}).  In these cases, we kept the science frame only if the DTTS flux values taken both before and after the frame duration were within $\pm1$~MAD.

Applying this frame selection on the science master cubes only accounted for weather variations within the science sequences, and does not address the aforementioned issue of different weather conditions between the flux and science frames altering the measured companion contrasts.  To address this, we also applied the DTTS flux value frame selection approach to the flux observations.  However, since the DTTS records the target star PSF as it appears behind the coronagraph, and since the target star PSF is dithered out from behind the coronagraph during the flux observations, no DTTS flux values are available during the flux sequences.  Therefore, we kept or excluded entire flux cubes based on the temporally-closest DTTS flux values.  

The resulting flux spectra of HD~112863~B and HD~206505~B can be seen in Fig.~\ref{fig:hd112863wDTTSnoCC} and Fig.~\ref{fig:hd206505DTTS}.  While the offset between the $H23$/$YJ$ and $K12$/$YJH$ spectra of HD~206505~B is now resolved, an offset remains between the $H23$/$YJ$ and $K12$/$YJH$ spectra of HD~112863~B.  This remaining offset is due to the attenuation of HD~112863~B itself by the coronagraph.

\begin{figure*}[h!]
    \centering
    \includegraphics[width=0.48\linewidth]{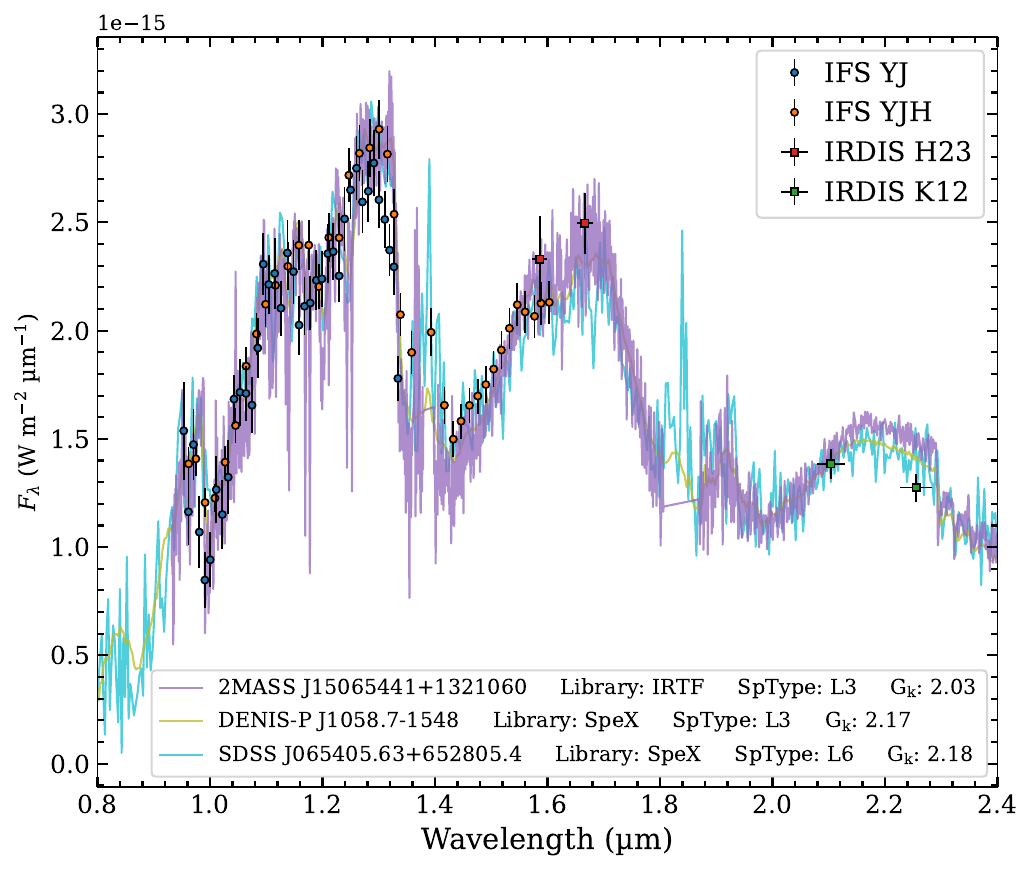}
    \includegraphics[width=0.48\linewidth]{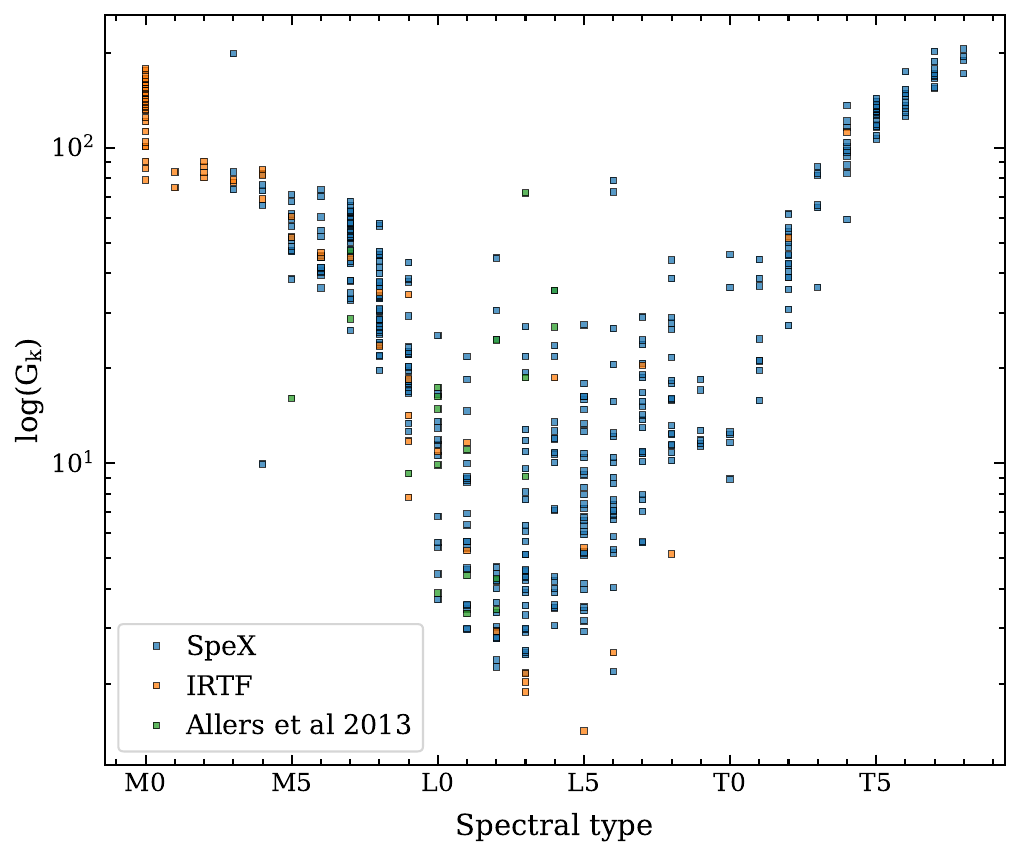}
    \caption{\textbf{Left:} The three best-fitting spectra to HD~112863~B, when compared to the spectra of M, L, and T dwarfs included in the SpeX, IRTF, and \cite{Allers2013} spectral libraries.  \textbf{Right:} The $G_{k}$ values from fitting each of the spectra of the M, L, and T dwarfs in the SpeX, IRTF, and \cite{Allers2013} spectral libraries to HD~112863~B.} 
    \label{fig:hd112863empfit}
\end{figure*}

\begin{figure*}[h!]
    \centering
    \includegraphics[width=0.48\linewidth]{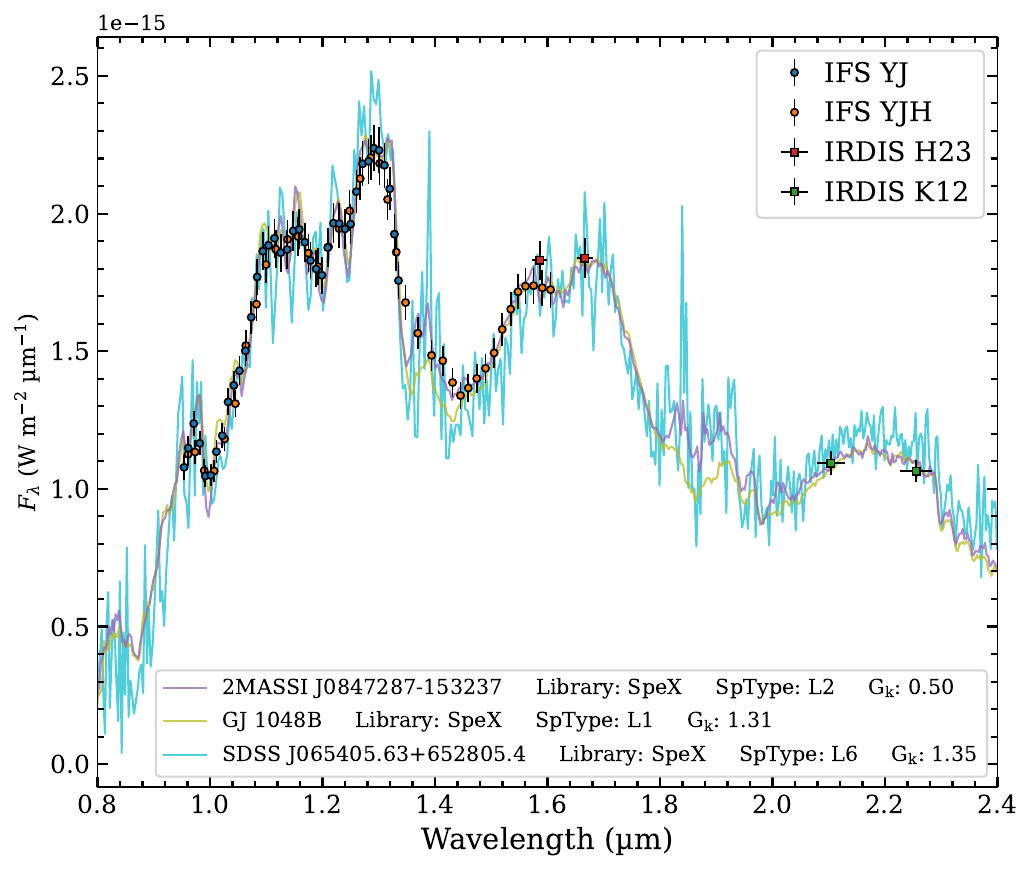}
    \includegraphics[width=0.48\linewidth]{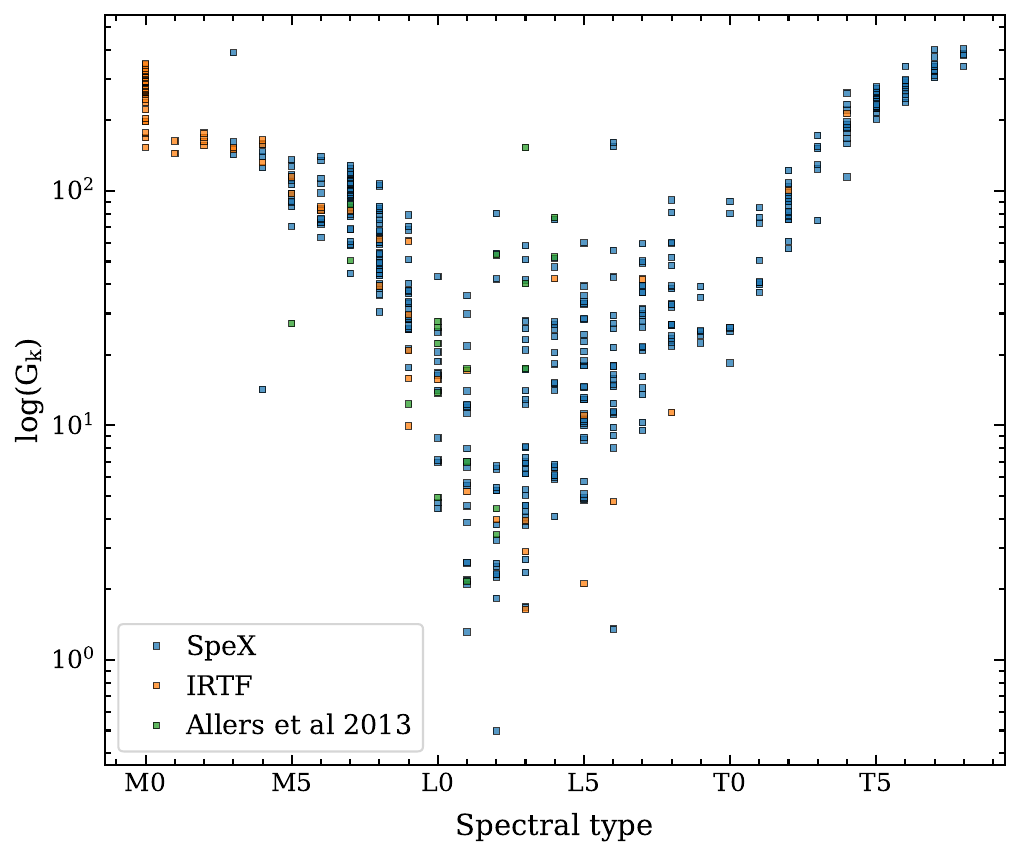}
    \caption{Same as Fig.~\ref{fig:hd112863empfit} but for HD~206505~B.}
    \label{fig:hd206505empfit}
\end{figure*}

\subsubsection{Coronagraph transmission correction} \label{subsubsec:coron}

HD~112863~B lies within the IWA ($\sim150$~mas) of the coronagraph in both the 2021-04-07 (at $\sim104$~mas) and 2022-01-30 (at $\sim149$~mas) observations, as already discussed in \cite{Rickman2024}, meaning that its flux is attenuated by the coronagraph in both datasets. To correct for this effect, we used the $H23$, $K12$, and $YJ$ coronagraphic transmission profiles of the N\_ALC\_YJH\_S coronagraph (see Sect.~\ref{sec:observs}), as measured on-sky (A. Vigan, see Fig.~\ref{fig:hd112863_coron_IRDIS} and Fig.~\ref{fig:hd112863_coron_YJ}).  Since the N\_ALC\_YJH\_S coronagraph uses a Lyot stop \citep{2019A&A...631A.155B}, and since a Lyot stop coronagraph is expected to not only attenuate the flux of an off-axis PSF, but also distort its shape (e.g. \citealp{Lloyd2005}), we corrected the science frames directly for the coronagraph transmission, in order to properly rectify both the astrometry and photometry of the companion.

We used cubic spline fits to the measured transmittance values to create radial coronagraphic transmission images, and then divided the science frames of HD~112863 by these images.  $YJH$ coronagraph transmission measurements were not available, so we used the $YJ$ and $H23$ coronagraph transmission measurements to perform a bivariate spline fit across both the separation $\rho$ and $\lambda$ axes, and then extracted the coronagraphic transmission profiles from this fit at each wavelength of IFS $YJH$ (see Fig.~\ref{fig:hd112863_coron_YJH}).  Since the spline fits to the measured values would occasionally include flux ratios $>1$, we set all values in the radial coronagraphic transmission images $>1$ to $1$, and also set all pixel values at $\rho>80$~pixels in IRDIS (and at $\rho > 40$~pixels in IFS) to $1$.  To avoid unnecessarily increasing the flux of the HD~112863 in the science images, to the point that it would disrupt TRAP's ability to estimate the background noise during post-processing, we set the pixels $\rho<1$~full width half maximum (FWHM) in the coronagraph transmission images to the flux ratio value at $\rho=1$~FWHM.  

The resulting flux spectra of HD~112863~B is shown in Fig.~\ref{fig:hd112863DTTSandCC}.  We also note that the approach used to correct for the coronagraphic attenuation is only a first-order approximation.  While our approach does appear to properly account for this attenuation (see Fig.~\ref{fig:hd112863wDTTSnoCC} versus Fig.~\ref{fig:hd112863DTTSandCC}), a fully consistent approach, that would include proper treatment of Fourier optics, would require generating and using a forward model of the companion PSF that includes the distortion and attenuation of the coronagraph applied on the flux frames in the Fourier domain, instead of applying coronagraph correction directly on the science frames.  However, generating such images is beyond the scope of this work.

\section{Spectral analysis}  \label{sec:allspectrafits}

\subsection{Empirical spectra comparisons} \label{subsec:empfits}

To estimate the spectral types of HD~112863~B and HD~206505~B, we compared the objects' spectra to that of the M, L, and T dwarf standards included in the SpeX \citep{Burgasser2014}, IRTF \citep{Cushing2005}, and \cite{Allers2013} spectral libraries.  The spectra included in these libraries cover wavelength ranges of at least $0.9$ to $2.4$~$\mu$m, and a range of resolutions~$\approx75$ to $2000$.  We note that the \cite{Allers2013} library is comprised of objects that are suspected to be $\lesssim300$~Myr old.

We used the \texttt{species} Python package \citep{Stolker2020} to compare each empirical spectra in these libraries to HD~112863~B and HD~206505~B.  \texttt{species} follows the approach described in \cite{Cushing2008} for determining the best-fitting empirical spectra, in which the goodness-of-fit statistic $G_k$ is defined as 

\begin{equation}
\label{eq:Gk}
G_k = \sum_{i=1}^n w_i \left ( \frac{f_i - C_k\mathcal{F}_{k,i}}{\sigma_i} 
\right)^2,
\end{equation}

\noindent where $n$ is the number of spectrophotometric data points, $w_i$ is the weight of spectophotometric data point $i$, $f_i$ is the flux density of spectrophotometric data point $i$, $\sigma_i$ is the error of spectrophotometric data point $i$, $\mathcal{F}_{k, i}$ is the flux density of the corresponding spectrophotometric point from the empirical spectrum $k$, and $C_k$ is the scaling parameter.  If $i$ is a photometric data point, $w_i$ is set to the FWHM of the filter associated with $i$, while $\mathcal{F}_{k, i}$ is the computed synthetic photometry of empirical spectra $k$ using the photometric filter associated with $i$.  If $i$ is a spectroscopic data point, $w_i$ is set to the spacing between wavelengths in the spectral data points, while $\mathcal{F}_{k, i}$ is computed by interpolating the empirical spectra $k$ to the wavelength of the spectral data associated with $i$.  The flux density of each empirical spectra is scaled to match the flux of the object via $C_k$, which is defined as

\begin{equation}
\label{eq:C}
C_k = \frac{\sum w_i f_i \mathcal{F}_{k,i} / \sigma_i^2}{\sum w_i 
  \mathcal{F}_{k,i}^2 / \sigma_i^2}.
\end{equation}

\noindent A more detailed description of Eq.~\ref{eq:Gk} and Eq.~\ref{eq:C} can be found in \cite{Cushing2008}.

The empirical spectra fitting module included in the latest version of \texttt{species} only calculates $G_k$ for an object based on a single spectrum. Calculating $G_k$ for an object with multiple spectra (such as both $YJ$ and $YJH$), along with photometry (such as $H23$ and $K12$), as is the case for HD~112863~B and HD~206505~B, required updating this module in \texttt{species} to include $i$ from multiple spectra, as well as $i$ corresponding to photometric points.  These changes also included the aforementioned treatment of $w_i$, since in the latest version of \texttt{species}, $w_i = 1.0$ for all $i$, with no consideration of weighting differences for spectroscopic versus photometric points.

We chose not to test different visual extinction $A_V$ values or different RV values when calculating $G_k$, since extinction is negligible for each target, and since the resolution of our observed spectra are too low to detect any RV shift.  Specifically, for HD~112863, $A_V \approx 0$~mag, while for HD~206505, $A_V \approx 0.0008$~mag, therefore indicating that any extinction or reddening effects are negligible for both targets\footnote{$A_V$ estimates taken from \textit{Gaia} data release 3 \citep{GaiaCollaboration2023}} \citep{Cardelli1989}.

The best-fitting empirical spectra for our targets, which we consider to be the three empirical spectra with the lowest $G_k$ among all objects included in SpeX, IRTF, and \cite{Allers2013}, are shown in Fig.~\ref{fig:hd112863empfit} (for HD~112863~B) and Fig.~\ref{fig:hd206505empfit} (for HD~206505~B).  We reviewed the literature to ensure that none of the three best-fitting objects have controversial spectral classifications, and that none of the objects are binaries; the original two best-fitting objects to HD~112863~B were binaries, so we excluded them from our analysis.  Given that the L$6$-type object SDSS~J065405.63+652805.4 does not match the minimum in $\log{(G_k)}$ vs spectral type of both Fig.~\ref{fig:hd112863empfit} and Fig.~\ref{fig:hd206505empfit}, we suspect that this object is misclassified or an unresolved binary, and therefore exclude it in our spectral typing of both companions.  Thus, the empirical spectra fits show that HD~112863~B corresponds to a L$3\pm1$ spectral type, while HD~206505~B corresponds to a L$2\pm1$ spectral type, confirming the results from our updated CMD (see Appendix~\ref{appendix_CMD}, specifically Fig.~\ref{fig:cmd}).  

\subsection{Atmospheric model fits} \label{subsec:atmofits}

\begin{figure}[t]
    \centering
    \includegraphics[width=0.5\textwidth]{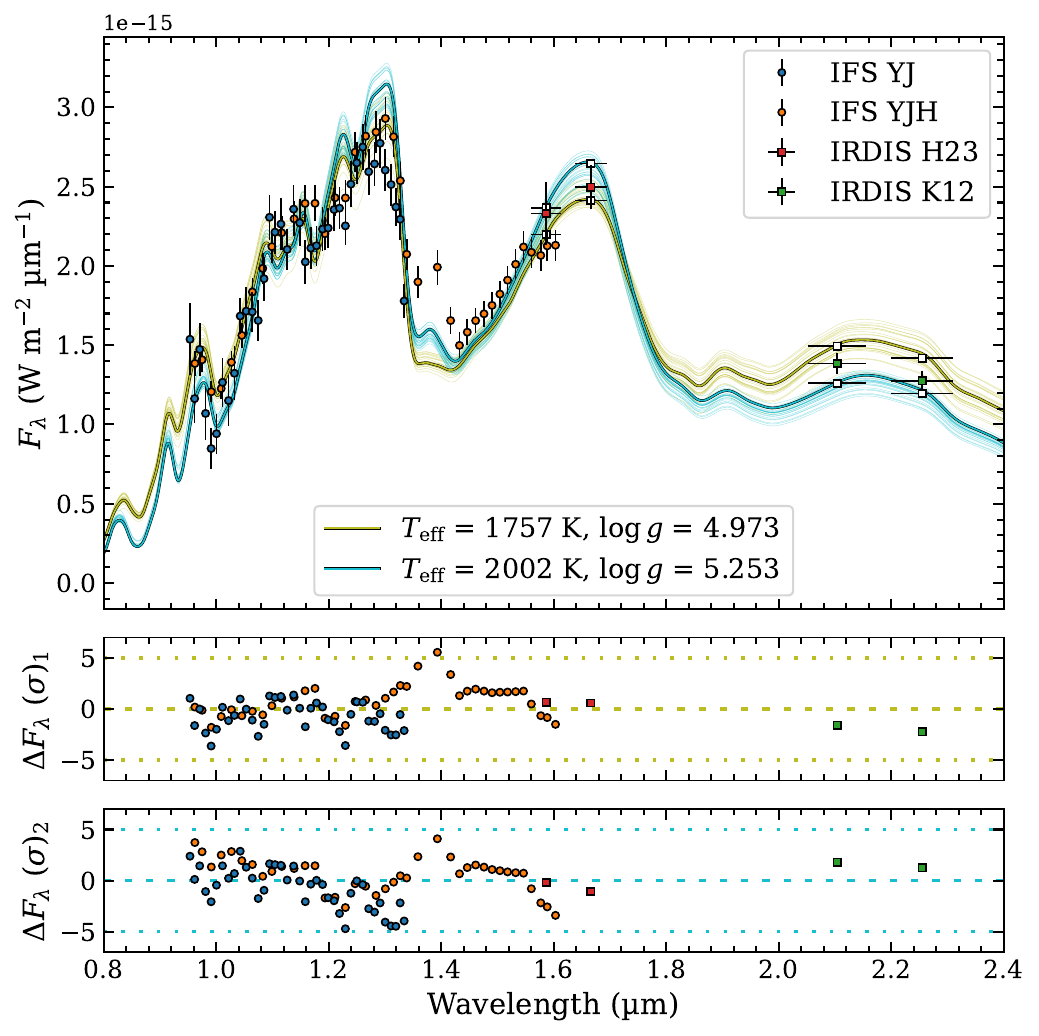}
    
    \caption{The median atmospheric model of the first mode (\textit{demarcated olive line}) and the median atmospheric model of the second mode (\textit{demarcated cyan line}) from the BT-Settl grid fit to the SPHERE photometry and spectroscopy of HD~112863~B (see Fig.~\ref{fig:hd112863_btsettl_posts}).  Included are $25$~models chosen randomly from the first mode (\textit{translucent olive lines}) and $25$~models chosen randomly from the second mode (\textit{translucent cyan lines}) of the posterior distribution.  The residuals between the SPHERE data and both the first mode ($\Delta F_\lambda (\sigma)_1$) as well as the second mode ($\Delta F_\lambda (\sigma)_2$) are shown.}
    \label{fig:hd112863_btsettl_main}
\end{figure}

To constrain the physical characteristics of HD~112863~B and HD~206505~B, we fit grids of atmospheric model spectra to the SPHERE spectra of each object.  We used the BT-Settl \citep{Allard2011,Allard2012} and the Sonora Diamondback \citep{Morley2024} atmospheric model grids, given their suitability for our targets.  In particular, we chose these models since they are applicable to both BDs and VLMSs; given that our objects are early-to-mid L dwarfs (see Sect.~\ref{subsec:empfits}) and have $M$ near the HBMM (see Appendix~\ref{appendix_orbitfits}), both objects could be either a BD or a VLMS, so any model grid used must accommodate both possibilities.  Additionally, we chose these grids because they include some form of cloud treatment, since cloud formation is relevant both for BDs (\citealp{Ackerman2001}, \citealp{Helling2008}, \citealp{Stephens2009}) and for UCDs as a whole \citep{Marley2002, Cushing2006, Cushing2008, Saumon2008}.

The BT-Settl atmospheric model grid was one of the first sets of simulated spectra to be able to match the L-T transition.  The BT-Settl grid is also able to match the M dwarf population in $T_{\rm{eff}}$ vs color space more accurately compared to previous model grids \citep{Allard2012}. The Sonora Diamondback atmospheric model grid consists of spectra that act as boundary conditions for new evolutionary models, which are calculated for objects ranging from $0.5$~$M_{\rm{Jup}}$ to $84$~$M_{\rm{Jup}}$.  Both metallicity $\rm{[M/H]}$ and sedimentation efficiency $f_{\rm{sed}}$ (inversely corresponding to cloud thickness) are included as varied parameters within the Sonora Diamondback atmospheric model grid, which is distinct compared to older and more simplistic grids where $T_{\rm{eff}}$ and $\log{g}$ are typically the only available ``free'' parameters \citep{Morley2024}.

\begin{figure}[t]
    \centering
    \includegraphics[width=0.5\textwidth]{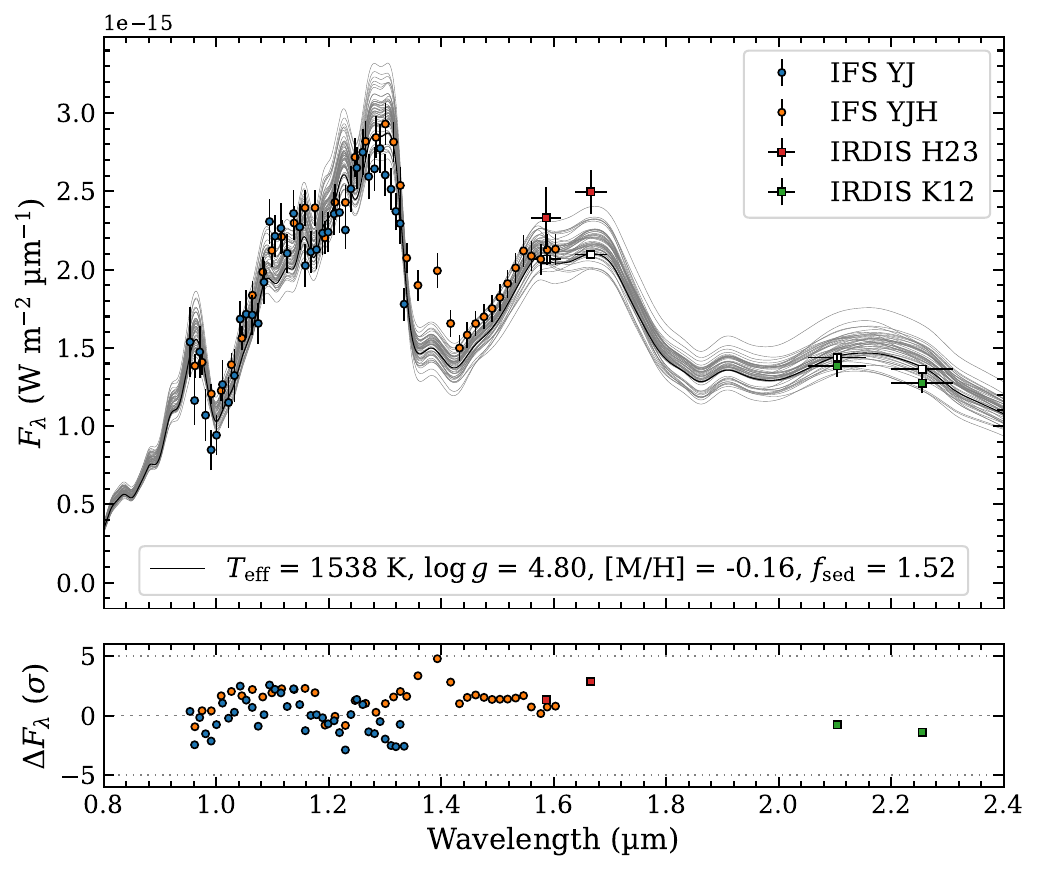}
    
    \caption{The median atmospheric model (\textit{black line}, see Fig.~\ref{fig:hd112863_sonoradiamond_posts}) from the Sonora Diamondback grid fit to the SPHERE photometry and spectroscopy of HD~112863~B, along with $50$~models chosen randomly from the posterior distribution (\textit{gray lines}).}
    \label{fig:hd112863_sonoradiamond_main}
\end{figure}

\begin{table}[h]
    \caption{Fitted and derived parameter values from the atmospheric model fits to HD~112863~B, along with the $\log{Z}$ associated with each fit (see Fig.~\ref{fig:hd112863_btsettl_posts} and Fig.~\ref{fig:hd112863_sonoradiamond_posts}).}
    \centering
    \begin{tabular}{ccc}
         \hline \hline
         Parameter & BT-Settl & Sonora Diamondback \\
         \hline \hline
         $T_{\rm{eff}}$~(K) & $1757^{+37}_{-36}, 2002^{+23}_{-24}$ & $1538^{+73}_{-75}$ \\
         $\log{g}$  & $4.973^{+0.057}_{-0.063}, 5.253^{+0.037}_{-0.033}$ & $4.80^{+0.10}_{-0.11}$ \\
         $\rm{[M/H]}$ & -- & $-0.16^{+0.26}_{-0.26}$ \\
         $f_{\rm{sed}}$  & -- & $1.52^{+0.71}_{-0.52}$ \\
         $R$\tablefoottext{a} ($R_{\rm{Jup}}$) & $1.428^{+0.082}_{-0.078}, 1.039^{+0.041}_{-0.059}$ & $1.74^{+0.22}_{-0.19}$ \\
         $\log{L/L_{\odot}}$ &  $-3.733^{+0.023}_{-0.027}, -3.790^{+0.020}_{-0.020}$ &  $-3.785^{+0.025}_{-0.025}$\\
         \hline 
         $\log{Z}$ & $2917.46\pm0.40$ & $2810.69\pm0.17$ \\
         \hline \\
    \end{tabular}
    \tablefoot{
    \tablefoottext{a}{For both model fits, $R$ is a derived parameter that is used to scale the flux of each spectra, along with $\varpi$.}}\label{tab:hd112863atmofitresults}
\end{table}

We fit the atmospheric models to each object using \texttt{species}.  \texttt{species} uses a given atmospheric model grid to fit physical parameters to an observed spectrum by linearly interpolating between the synthetic spectra within the grid, and extrapolating from the interpolated spectrum to obtain the corresponding parameter values.  To probe the posterior distribution of each physical parameter, we used nested sampling with \texttt{species} via the \texttt{Ultranest} Python package \citep{Buchner2016, Buchner2019, Buchner2021}; nested sampling is better at probing multi-model distributions, that occur when fitting atmospheric model grids, compared to Markov chain Monte Carlo (MCMC) algorithms (e.g. \citealp{Hurt2024}).  We sampled the posterior distributions of each model grid's parameters using $500$~live points.  Using gaussian kernel density estimation, we computed the highest posterior density interval for each parameter, in order to properly estimate the median and $1\sigma$ intervals for any posteriors that may be multi-modal.  For any fits with multi-modal posteriors, we compared the samples from a single mode of one parameter with the those of all other multi-modal parameters, to correctly match the corresponding modes between all parameters.

\begin{figure}[h]
    \centering
    \includegraphics[width=0.5\textwidth]{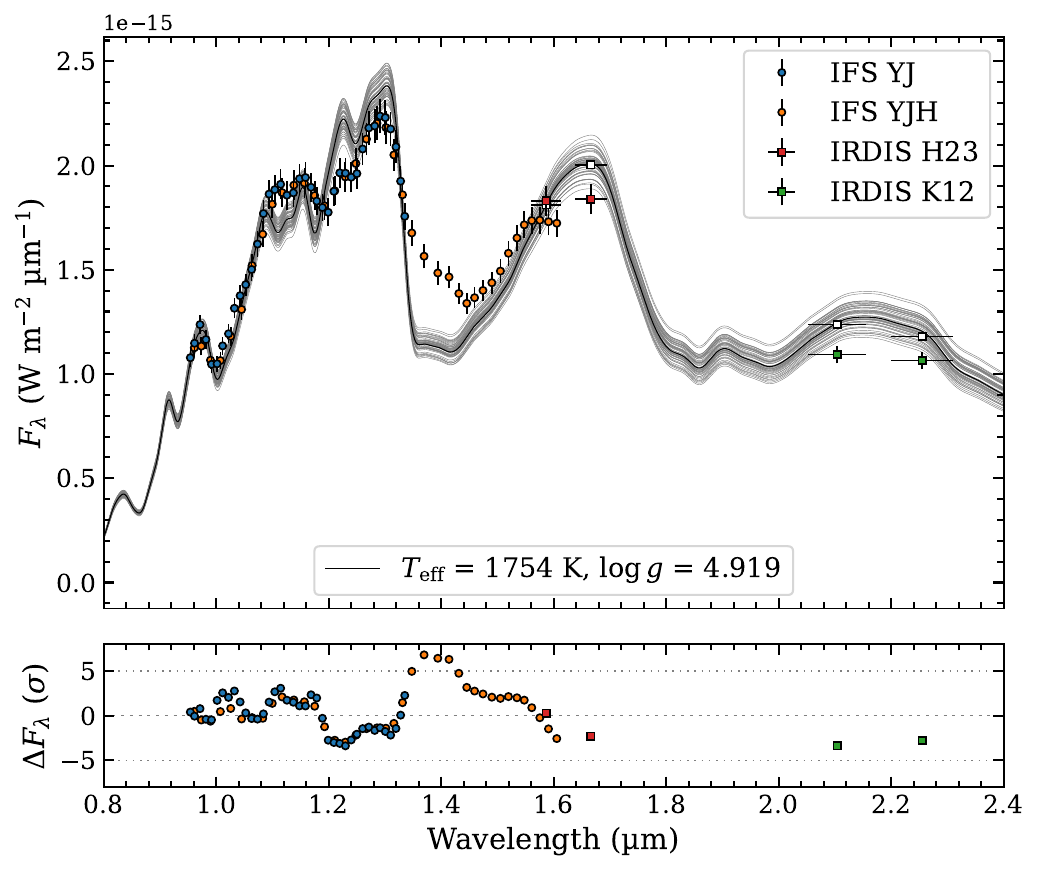}
    
    \caption{The median atmospheric model (\textit{black line}, see Fig.~\ref{fig:hd206505_btsettl_posts}) from the BT-Settl grid fit to the SPHERE photometry and spectroscopy of HD~206505~B, along with $50$~models chosen randomly from the posterior distribution (\textit{gray lines}).}
    \label{fig:hd206505_btsettl_main}
\end{figure}

When fitting each of the atmospheric model grids, we included a Gaussian prior on the $M$ of HD~112863~B and HD~206505~B, using their dynamical masses from the updated orbit fits ($M = {77.7}_{-3.3}^{+3.4}$~$M_{\rm{Jup}}$ and $M = 79.8\pm1.8$~$M_{\rm{Jup}}$, respectively, see Appendix~\ref{appendix_orbitfits}).  We also included a Gaussian prior on the $\varpi$ of each target ($\varpi = 26.96 \pm 0.03$~mas and $\varpi = 22.77 \pm 0.02$~mas, respectively, \citealp{Rickman2024}).  We set a uniform prior on the effective temperature $T_{\rm{eff}}$ with lower and upper bounds of $1300$~K to $2500$~K, based on previous observational studies of early-to-mid L dwarfs (\citealp{Kirkpatrick2005}, \citealp{Rajpurohit2012}, \citealp{Gizis2013}, \citealp{Dieterich2014}, \citealp{Lodieu2014}, \citealp{Filippazzo2015}, \citealp{Dupuy2017}, \citealp{Hurt2024}, \citealp{Li2024}).  \texttt{species} uses the $\log{g}$ and $R$ value at each step to calculate the corresponding $M$, and thereby ensures that the prior on $M$ is respected when computing the log-likelihood.  $L$ is a derived parameter calculated using the posteriors of $T_{\rm{eff}}$ and $R$.  

As done for the empirical spectra fitting (see Sect.~\ref{subsec:empfits}), we did not include any parameters to account for any potential RV shift, $A_V$, or visual reddening $R_V$, when computing the atmospheric model fits.  Moreover, we did not include any parameters in our fits to account for rotational broadening, since the resolution of our observed spectra are too low to detect this phenomenon.  We also chose to apply weights to each observed spectrophotometric point when calculating the log-likelihood, following the same approach as described in Sect.~\ref{subsec:empfits}: photometric points are weighted based on the FWHM of the associated filter, while spectroscopic points are weighted based on the spacing between wavelengths. 

\begin{figure}[h]
    \centering
    \includegraphics[width=0.5\textwidth]{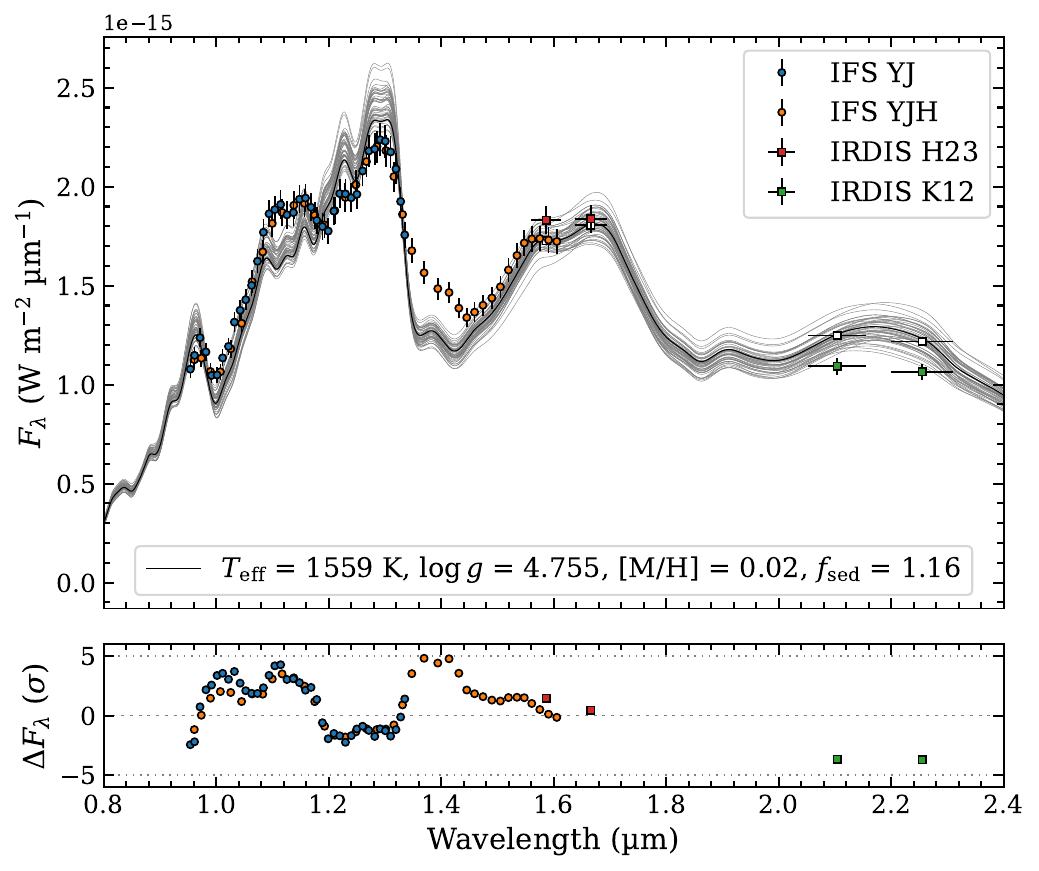}
    \caption{The median atmospheric model (\textit{black line}, see Fig.~\ref{fig:hd206505_sonoradiamond_posts}) from the Sonora Diamondback grid fit to the SPHERE photometry and spectroscopy of HD~206505~B, along with $50$~models chosen randomly from the posterior distribution (\textit{gray lines}).}
    \label{fig:hd206505_sonoradiamond_main}
\end{figure}

\begin{table}[h]
    \caption{Same as Table~\ref{tab:hd112863atmofitresults} but for HD~206505~B (see Fig.~\ref{fig:hd206505_btsettl_posts} and Fig.~\ref{fig:hd206505_sonoradiamond_posts}).}
    \centering
    \begin{tabular}{ccc}
         \hline \hline
         Parameter & BT-Settl & Sonora Diamondback \\
         \hline \hline
         $T_{\rm{eff}}$~(K) & $1754^{+13}_{-13}$  & $1559^{+43}_{-43}$ \\
         $\log{g}$  & $4.919^{+0.031}_{-0.029}$ & $4.755^{+0.065}_{-0.065}$ \\
         $\rm{[M/H]}$ & -- & $0.02^{+0.22}_{-0.22}$ \\
         $f_{\rm{sed}}$  & -- & $1.16^{+0.28}_{-0.16}$ \\
         $R$\tablefoottext{a} ($R_{\rm{Jup}}$) & $1.543^{+0.057}_{-0.053}$ & $1.86^{+0.14}_{-0.13}$ \\
         $\log{L/L_{\odot}}$ & $-3.669^{+0.019}_{-0.021}$ &  $-3.709^{+0.019}_{-0.021}$ \\
         \hline 
         $\log{Z}$ & $2963.52\pm0.21$ & $2854.15\pm0.11$ \\
         \hline  \\
    \end{tabular}
    \tablefoot{
    \tablefoottext{a}{For both model fits, $R$ is a derived parameter that is used to scale the flux of each spectra, along with $\varpi$.}}
    \label{tab:hd206505atmofitresults}
\end{table}

The atmospheric model fit results for HD~112863~B are shown in Fig.~\ref{fig:hd112863_btsettl_main} using the BT-Settl grid and in Fig.~\ref{fig:hd112863_sonoradiamond_main} using the Sonora Diamondback grid.  The parameter values from both of these fits, along with the evidence of each model $Z$, are compared in Table~\ref{tab:hd112863atmofitresults}.  The BT-Settl fit is strongly bimodal (see Fig.~\ref{fig:hd112863_btsettl_main} and Fig.~\ref{fig:hd112863_btsettl_posts}), although the first mode (corresponding to $T_{\rm{eff}}=1757^{+37}_{-36}$~K, $\log{g}=4.973^{+0.057}_{-0.063}$, etc.) is $\approx 3$~times higher in probability for $T_{\rm{eff}}$ compared to the second mode (corresponding to $T_{\rm{eff}}=2002^{+23}_{-24}$~K, $\log{g}=5.253^{+0.037}_{-0.033}$, etc.), indicating that the first mode is somewhat preferred.  One possible source of this bimodality is an exchange between fitting to the peak at $1.4$~$\mu$m and fitting to the shape of the spectra from $1.2-1.3$~$\mu$m.  Fig.~$1$ of \cite{Hurt2024} shows that BT-Settl models of higher $T_{\rm{eff}}$, and somewhat similarly, those of higher $\log{g}$, have higher flux at $1.4$~$\mu$m, which better matches the peak at $1.4$~$\mu$m in the spectra of HD~112863~B.  Meanwhile, BT-Settl models of lower $T_{\rm{eff}}$ show a lower flux around $1.2-1.3$~$\mu$m compared to those of higher $T_{\rm{eff}}$, better matching the spectra of HD~112863~B at $1.2-1.3$~$\mu$m.

The results from both fits indicate that the $T_{\rm{eff}}$ of HD~112863~B corresponds to an early-to-mid L spectral type (\citealp{Dieterich2014}, \citealp{Filippazzo2015}), which supports the conclusions made in Sect.~\ref{subsec:empfits}.  The results from both fits also indicate that HD~112863~B has a high $\log{g}$ ($\approx5$~dex, \citealp{Chabrier2000}), a characteristic of older UCDs \citep{Martin2017}, which coincides with the suspected older age ($>1$~Gyr) of the HD~112863 system \citep{Rickman2024}.  The results from the Sonora Diamondback fit indicate that HD~112863~B has thick cloud cover (since $f_{\rm{sed}}<2$, \citealp{Morley2024}), as well as a near-solar $\rm{[M/H]}$ (see Fig.~\ref{fig:hd112863_sonoradiamond_posts}).

The atmospheric model fit results for HD~206505~B are shown in Fig.~\ref{fig:hd206505_btsettl_main} using the BT-Settl grid and in Fig.~\ref{fig:hd206505_sonoradiamond_main} using the Sonora Diamondback grid.  The parameter values from both of these fits, along with the $Z$ of each model, are compared in Table~\ref{tab:hd206505atmofitresults}.  Both fits for HD~206505~B indicate $T_{\rm{eff}}$ values approximately corresponding to an early-to-mid L type (\citealp{Dieterich2014}, \citealp{Filippazzo2015}), again supporting the conclusions from Sect.~\ref{subsec:empfits}.  The $\log{g}$ values from both fits are also high, therefore coinciding with the suspected older age of HD~206505~B \citep{Rickman2024}.  Finally, the Sonora Diamondback fit shows that HD~206505~B has thick cloud cover and that $\rm{[M/H]}$ is near-solar.

For both HD~112863~B and HD~206505~B, BT-Settl is the favored model grid over Sonora Diamondback, since the Bayes factor is $>100$ between the two models for both objects, respectively (see Table~\ref{tab:hd112863atmofitresults} and Table~\ref{tab:hd206505atmofitresults}).  We note that the $\rm{[M/H]}$ found for both HD~112863~B and HD~206505~B from the Sonora Diamondback fits (see Table~\ref{tab:hd112863atmofitresults} and Table~\ref{tab:hd206505atmofitresults}) are consistent with $\rm{[Fe/H]}$ found for the host stars in \cite{Rickman2024}.

While $R$ measurements from transits of UCDs with measured masses near the HBMM find $0.8$~$R_{\rm{Jup}}\lesssim R \lesssim
 1.5$~$R_{\rm{Jup}}$ (\citealp{Pont2006}, \citealp{Canas2018}, \citealp{Acton2020}, \citealp{Grieves2021}, \citealp{Canas2022}, \citealp{Sebastian2022}, \citealp{Lambert2023}, \citealp{FerreiradosSantos2024}), the results from nearly all of our atmospheric model fits yield $R$ near or above the upper bound of this range, with the only exception being the first mode of the BT-Settl fit to HD~112863~B.  Discrepancies regarding $R$ computed from atmospheric modeling are a known issue (e.g. \citealp{Zalesky2019}, \citealp{Lueber2022}, \citealp{Hood2023}, \citealp{Zhang2023}, \citealp{Tobin2024}), although this usually involves an underestimation of $R$ by the models, not an overestimation.  One potential explanation for this may lie in the estimate of $\log{g}$ found by our model fits; $\log{g}$ for older ($>1$~Gyr) UCDs specifically near the HBL are expected to be $\approx5.5$~dex \citep{vonBoetticher2019}, higher than our estimates, which would correspond to a lower $R$.  Another possible explanation for the unphysically large $R$ in our atmospheric model fits is that HD~112863~B and/or HD~206505~B binaries.  However, we note that neither HD~206505~B nor HD~112863~B demonstrate signs of binarity in their PSF shapes in the SPHERE data; from the SPHERE images, binarity is ruled out for both companions down to approximately $2$~AU.

\begin{figure}[h]
    \centering
    \begin{subfigure}[t]{0.5\textwidth}
    \includegraphics[width=\textwidth]{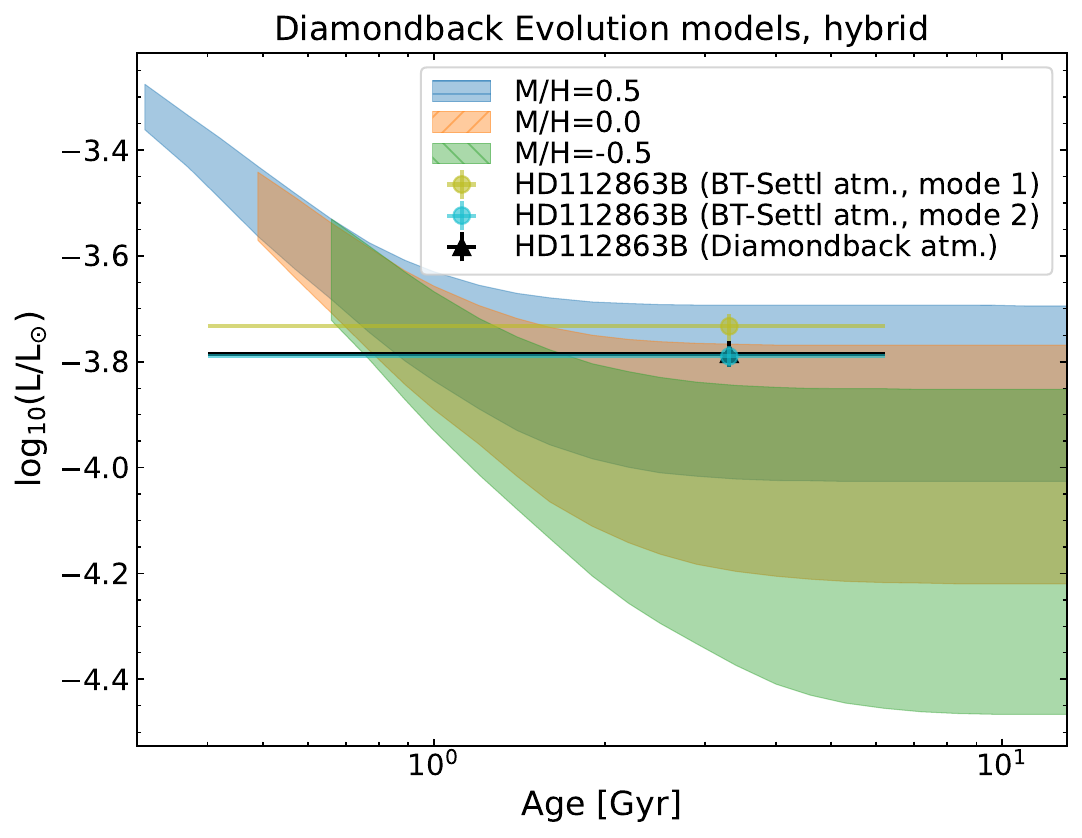} 
    \caption{} \label{fig:hd112863_evol_hybrid}
    \end{subfigure}
    \begin{subfigure}[t]{0.5\textwidth}
    \includegraphics[width=\textwidth]{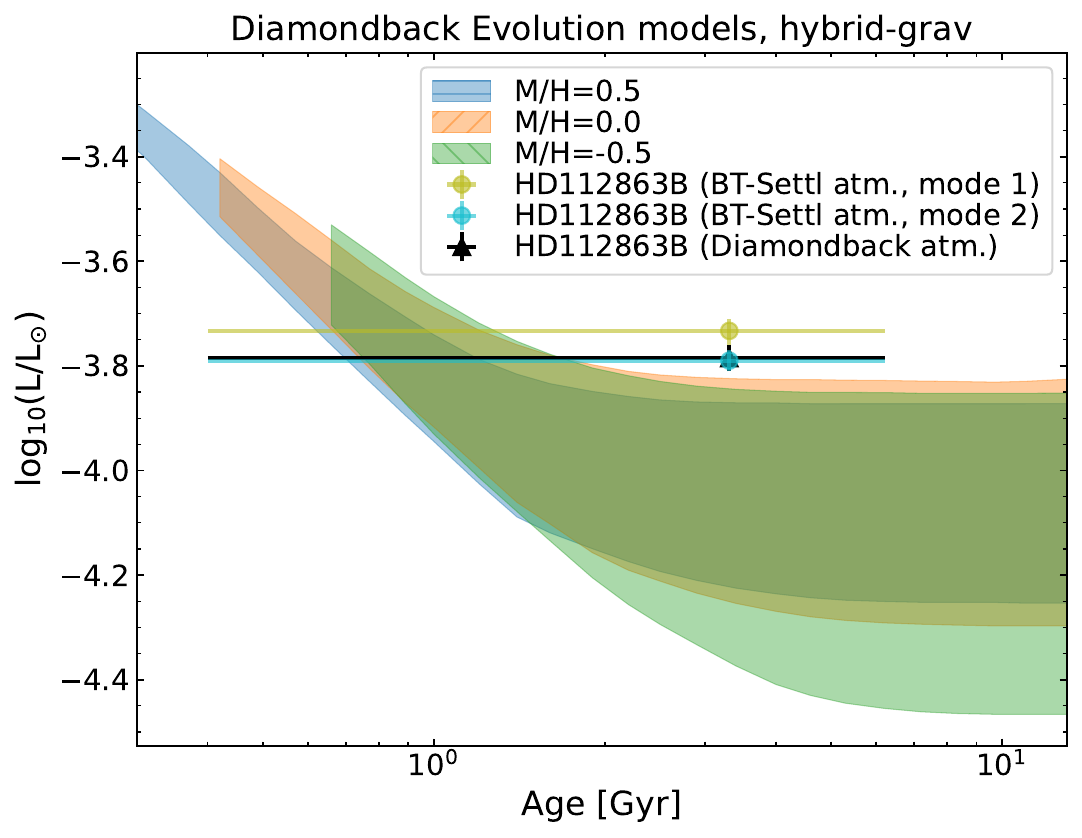}
    \caption{} \label{fig:hd112863_evol_hybrid_grav}
    \end{subfigure}
    \caption{The $L$ and system age of HD~112863~B compared to the Sonora Diamondback ``Hybrid'' \textbf{(a)} and ``Hybrid-grav'' \textbf{(b)} evolutionary models \citep{Morley2024}.  The $L$ values of both the first mode (\textit{olive circle}) and the second mode (\textit{cyan circle}) from the BT-Settl atmospheric model fit to HD~112863~B are included (see Fig.~\ref{fig:hd112863_btsettl_posts}).  The shaded regions show the model $L$ predictions that correspond to the $1\sigma$ dynamical mass constraint of HD~112863~B (${77.7}_{-3.3}^{+3.4}$~$M_{\rm{Jup}}$, see Appendix~\ref{appendix_orbitfits}), with different colors corresponding to one of the three $\rm{[M/H]}$ values used by the models.} \label{fig:hd112863_evol_both}
\end{figure}

 \cite{Hurt2024} found that proper treatment for modeling dust grains in the atmospheres of UCDs may explain model discrepancies, including at $1.4$~$\mu$m, where our residuals are typically largest ($\sim5\sigma$, see Fig.~\ref{fig:hd112863_btsettl_main}, \ref{fig:hd112863_sonoradiamond_main}, \ref{fig:hd206505_btsettl_main}, and \ref{fig:hd206505_sonoradiamond_main}).  \cite{Hurt2024} also found that dust grains may also explain the lower-than-expected $\log{g}$ values, and conversely the higher-than-expected $R$ values, found in some UCD atmospheric model fits.  Additionally, the large residuals around $1.4$~$\mu$m in our model fits may be the result of an imperfect telluric correction, which can occur during observations with variable atmospheric conditions where simultaneous monitoring of the host star's flux is not available, as described in \cite{Brown-Sevilla2023}.  However, the shape of the spectra of HD~112863~B and HD~206505~B, including around $1.4$~$\mu$m, match that of other objects, per our comparisons with empirical spectra (see Sect.~\ref{subsec:empfits}).  This implies that the large residuals in our atmospheric model fits are not due to any issues unique to our observations or objects' spectra, but instead may be the result of limitations in the models.  Although one such limitation may be incomplete water vapor opacity line lists included in the models, particularly in the older BT-Settl grids, this is unlikely to explain the large residuals seen at $1.4$~$\mu$m.  Both sets of models fail to fit this region of the spectra, including Sonora Diamondback, which includes the latest water line lists.  Overall, a thorough investigation of the tendency towards unphysically large $R$ in our model fits is beyond the scope of this work.

\section{Evolutionary model tracks} \label{sec:evolmodel}

To investigate whether HD~112863~B and HD~206505~B are stellar or substellar, and how they compare to evolutionary tracks, we compared the atmospheric-model-derived $L$ of both companions with $L$ versus age evolutionary models of UCDs.  We chose to use the Sonora Diamondback evolutionary models for this purpose, since they include the evolution of both VLMSs and BDs with a range of characteristics.  More specifically, the Sonora Diamondback evolutionary models include tracks without clouds, clouds dependent on $T_{\rm{eff}}$ (``Hybrid''), and clouds dependent on $T_{\rm{eff}}$ and $\log{g}$ (``Hybrid-grav''), as well as subsolar ($-0.5$), solar ($0.0$), and supersolar ($+0.5$) $\rm{[M/H]}$ \citep{Morley2024}.

Because clouds are relevant for modeling L dwarfs (see Sect.\ref{sec:intro} and Sect.~\ref{subsec:atmofits}), we chose to use only the two sets of Sonora Diamondback tracks that include clouds, which are the ``Hybrid'' and ``Hybrid-grav'' models.   We used the corresponding system ages listed in \cite{Rickman2024} as the age of each companion ($3.31 \pm 2.91$~Gyr for HD~112863~B and $3.94 \pm 2.51$~Gyr for HD~206505~B). 

\begin{figure}[h]
    \centering
    \begin{subfigure}[t]{0.5\textwidth}
    \includegraphics[width=\textwidth]{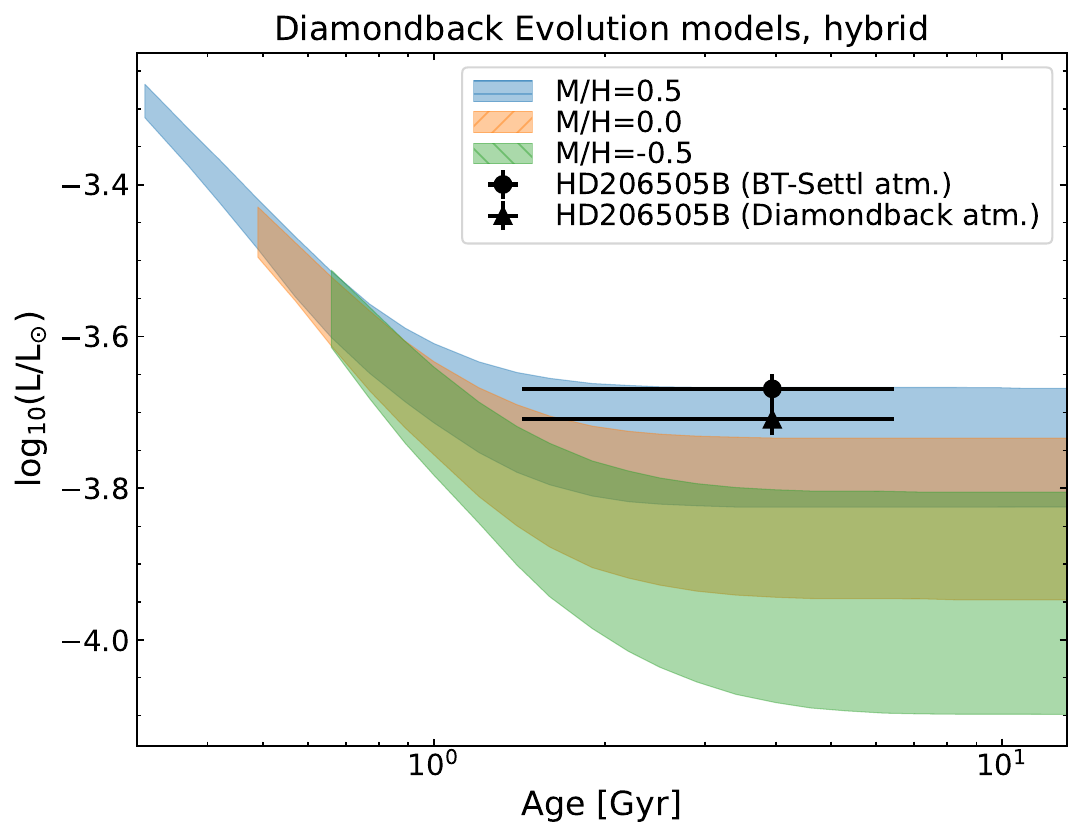} 
    \caption{} \label{fig:hd206505_evol_hybrid}
    \end{subfigure}
    \begin{subfigure}[t]{0.5\textwidth}
    \includegraphics[width=\textwidth]{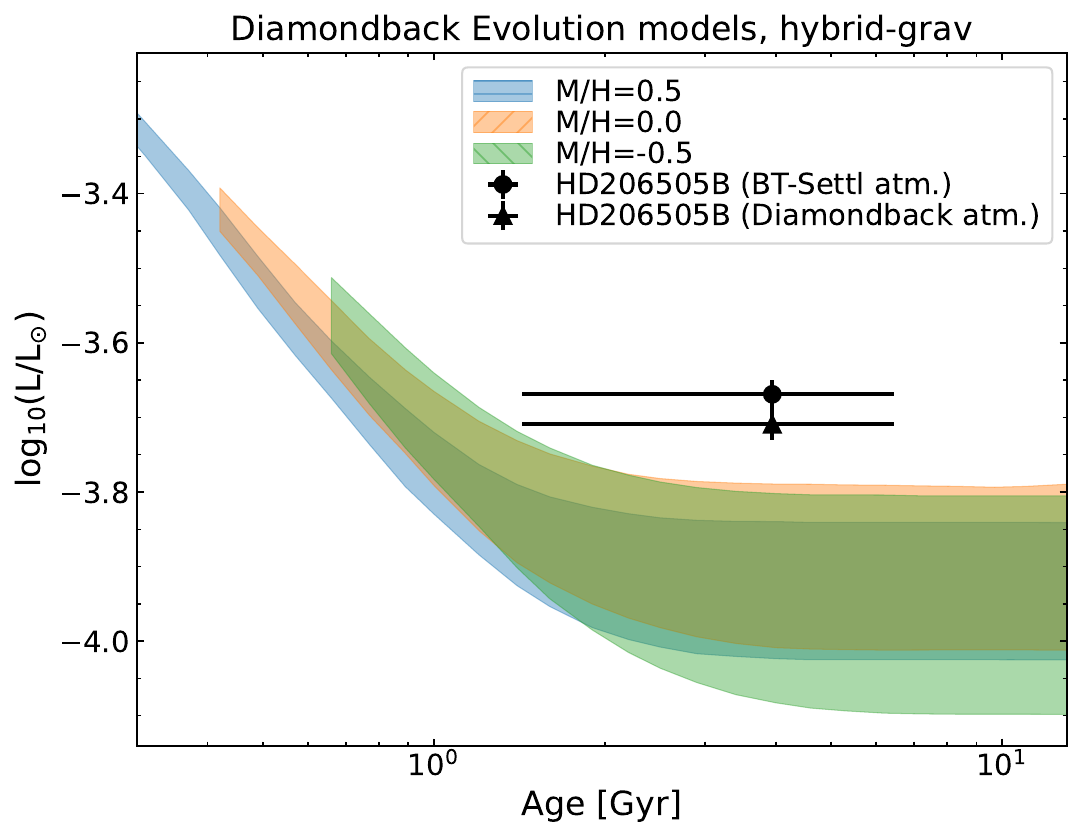}
    \caption{} \label{fig:hd206505_evol_hybrid_grav}
    \end{subfigure}
    \caption{Same as Fig.~\ref{fig:hd112863_evol_both} but for HD~206505~B.  The shaded regions show the model $L$ predictions that correspond to the $1\sigma$ dynamical mass constraint of HD~206505~B ($79.8\pm1.8$~$M_{\rm{Jup}}$, see Appendix~\ref{appendix_orbitfits}).} \label{fig:hd206505_evol_both}
\end{figure}

The comparison of HD~112863~B and HD~206505~B with the Sonora Diamondback $L$ evolutionary tracks are shown in Fig.~\ref{fig:hd112863_evol_both} and Fig.~\ref{fig:hd206505_evol_both}.  We find that the dynamical mass, measured $L$, and system ages are consistent with the evolutionary tracks for both objects for the ``Hybrid'' cases (see Fig.~\ref{fig:hd112863_evol_hybrid} and Fig.~\ref{fig:hd206505_evol_hybrid}, respectively).  From these comparisons, we also conclude that both HD~112863~B and HD~206505~B are likely VLMSs, given that $L$ in all of the tracks for both companions only correspond to objects that either have already reached a stable $L$ thanks to H burning, or will do so at some point as they age.

While the evolutionary model tracks are consistent with the companion $L$ and ages in the ``Hybrid'' scenario, there is a discrepancy between the companion $L$ and ages in the ``Hybrid-grav'' scenarios (see Fig.~\ref{fig:hd112863_evol_hybrid_grav} and Fig.~\ref{fig:hd206505_evol_hybrid_grav}), with the most tension occurring between ``Hybrid-grav'' tracks and the HD~206505~B BT-Settl $L$.  This is because the ``Hybrid-grav'' tracks predict a lower $L$ than the ``Hybrid'' tracks after $\approx1$~Gyr.  We note that this difference in $L$ beyond $1$~Gyr between the ``Hybrid'' and ``Hybrid-grav'' tracks is expected, since \cite{Morley2024} shows that ``Hybrid-grav'' models are $\approx100$~K cooler in $T_{\rm{eff}}$ beyond $1$~Gyr for objects $M\sim80$~$M_{\rm{Jup}}$ compared to ``Hybrid'' models.

\section{Conclusions}
\label{sec:conclus}

We report the spectral analysis of two benchmark L dwarfs orbiting HD~112863 and HD~206505.  From our results, we find that:

\begin{itemize}
    \item HD~112863~B is spectral type L$3\pm1$ while HD~206505~B is spectral type L$2\pm1$, confirming the conclusions from \cite{Rickman2024},
    
    \item for HD~112863~B, $T_{\rm{eff}}=1757^{+37}_{-36}$~K or $2002^{+23}_{-24}$~K, $\log{g}=4.973^{+0.057}_{-0.063}$ or $5.253^{+0.037}_{-0.033}$, $R=1.428^{+0.082}_{-0.078}$~$R_{\rm{Jup}}$ or $1.039^{+0.041}_{-0.059}$~$R_{\rm{Jup}}$, and $\log{L/L_{\odot}}=-3.733^{+0.023}_{-0.027}$ or $-3.790^{+0.020}_{-0.020}$,
    
    \item for HD~206505~B,  $T_{\rm{eff}}=1754^{+13}_{-13}$~K, $\log{g}=4.919^{+0.031}_{-0.029}$, $R=1.543^{+0.057}_{-0.053}$~$R_{\rm{Jup}}$, and $\log{L/L_{\odot}}= -3.669^{+0.019}_{-0.021}$,

    \item the $R$ found for both HD~112863~B and HD~206505~B in the majority of our atmospheric model fits are unphysically large, although this may be due to limitations with the atmospheric models themselves,
    
    \item the measured $L$, dynamical masses, and system ages of HD~112863~B and HD~206505~B are in agreement with the predictions from evolutionary models,
    
    \item both HD~112863~B and HD~206505~B are likely VLMSs.
\end{itemize}

The most significant difference in results compared to \cite{Rickman2024} is the photometry of each companion, as shown in Table~\ref{tab:updtd_astro_photo} (versus Table~$2$ of \citealp{Rickman2024}).  The biggest impact of this change is the CMD, where the placement of each companion, especially HD~112863~B, is more accurate (see Fig.~\ref{fig:cmd}).  Regardless, the overall spectral classification of HD~112863~B and HD~206505~B remain in agreement with this work.

We note that the constraints on the spectral and physical parameters of HD~206505~B are more precise than that of HD~112863~B (see Fig.~\ref{fig:hd206505empfit} versus Fig.\ref{fig:hd112863empfit}, Table~\ref{tab:hd206505atmofitresults} versus Table~\ref{tab:hd112863atmofitresults}, and Table~\ref{tab:orbitparameters}), a result of HD~112863~B being detected at a SNR~$\approx \times10$ lower than HD~206505~B (see Sect.~\ref{subsec:spectraextract}).  This is explained by the attenuation of HD~112863~B's flux by the coronagraph, in combination with the fact that HD~112863~B lies at a closer $\rho$ than HD~206505~B, in a region of higher noise.

HD~112863~B is one of a few companions successfully detected in HCI despite being within the IWA of an instrument's coronagraph (e.g. \citealp{Franson2024}).  Such detections aid in understanding the feasibility of observing companions in HCI at $\rho$ previously viewed as inaccessible by current instruments. 

The system ages are the limiting factor in determining where HD~112863~B and HD~206505~B are in their thermal evolution (as seen in Fig.~\ref{fig:hd112863_evol_both} and Fig.~\ref{fig:hd206505_evol_both}).  Future work on the HD~112863 and HD~206505 systems could therefore involve deriving stronger constraints on the system ages through asteroseismology (e.g. \citealp{Huber2009}, \citealp{Chaplin2011}).  Currently planned future work on these systems including further spectral analysis of both companions using data from VLTI/GRAVITY and VLT/HiRISE (Rickman et al., in prep.), which will provide constraints on the $\rm{C/O}$ ratio of the companions \citep{GRAVITYCollaboration2017}, and enable the detection of rotational broadening in their spectra \citep{Vigan2024}.  Finally, the additional spectra of HD~112863~B and HD~206505~B from these two sets of observations will provide verification of our conclusions presented here, including the tentative classification of both companions as VLMSs. 

 HD~112863~B and HD~206505~B add to the growing list of UCD and exoplanet companions detected in HCI with dynamical mass measurements.  Similar companions included in this list are HD~72946~B \citep[$M=69.5\pm0.5$~$M_{\rm{Jup}}$,][]{Balmer2023}, HD~4747~B \citep[$M=65.3^{+4.4}_{-3.3}$~$M_{\rm{Jup}}$,][]{Crepp2018}, HR~7672~B \citep[$M=72.7\pm0.8$~$M_{\rm{Jup}}$,][]{Brandt2019}, and HD~984~B \citep[$M=61\pm4$~$M_{\rm{Jup}}$,][]{Franson2022}.  Each of these are substellar companions are near the HBL and have $T_{\rm{eff}} \geq 1500$~K and $\log{g} \geq 5$, with HD~72946~B, HD~4747~B, and HR~7672~B specifically having ages $>1$~Gyr (\citealp{Balmer2023}, \citealp{Crepp2018}, \citealp{Meshkat2015}, \citealp{Wang2022}).  Building a larger sample of such objects is valuable, since they can be used collectively to empirically constrain mass-dependent boundaries such as the HBL (e.g. \citealp{Dupuy2017}), as well as more robustly test atmospheric and evolutionary models (e.g. \citealp{Grieves2021}, \citealp{Iyer2023}).  In particular, our work shows that current atmospheric models may be limited in explaining the full shape of spectra of older UCDs near the HBL, and in estimating their $R$; additional detections are needed to determine if these discrepancies are a true shortcoming of the models, or only unique to these two objects.

\begin{acknowledgements} We thank Jo\~ao Faria for his guidance regarding proper estimation of credible intervals for multi-modal posterior distributions.  We thank Matthias Samland for his insightful discussions that led to including weather-driven frame selection as a component of our data reduction procedure.  This work has been carried out within the framework of the National Centre of Competence in Research PlanetS supported by the Swiss National Science Foundation under grants 51NF40\_182901 and 51NF40\_205606. The authors acknowledge the financial support of the SNSF.  This publications makes use of the The Data and Analysis Center for Exoplanets (DACE), which is a facility based at the University of Geneva (CH) dedicated to extrasolar planets data visualization, exchange and analysis. DACE is a platform of the Swiss National Centre of Competence in Research (NCCR) PlanetS, federating the Swiss expertise in Exoplanet research. The DACE platform is available at \url{https://dace.unige.ch}. This work has made use of data from the European Space Agency (ESA) mission Gaia (\url{https://www.cosmos.esa.int/gaia}), processed by the Gaia Data Processing and Analysis Consortium (DPAC, \url{https://www.cosmos.esa.int/web/gaia/dpac/consortium}). Funding for the DPAC has been provided by national institutions, in particular the institutions participating in the Gaia Multilateral Agreement. This research made use of the SIMBAD database and the VizieR Catalog access tool, both operated at the CDS, Strasbourg, France. The original descriptions of the SIMBAD and VizieR services were published in \cite{2000A&AS..143....9W} and \cite{2000A&AS..143...23O}. This research has made use of NASA’s Astrophysics Data System Bibliographic Services.
\end{acknowledgements}

\bibliography{bib.bib} 

\begin{appendix}

\onecolumn
\section{Uncorrected and corrected spectra} \label{appendix_uncorrspecs}

\begin{figure*}[!h]
    \centering
    \begin{subfigure}[t]{0.49\linewidth}
    \includegraphics[width=\textwidth]{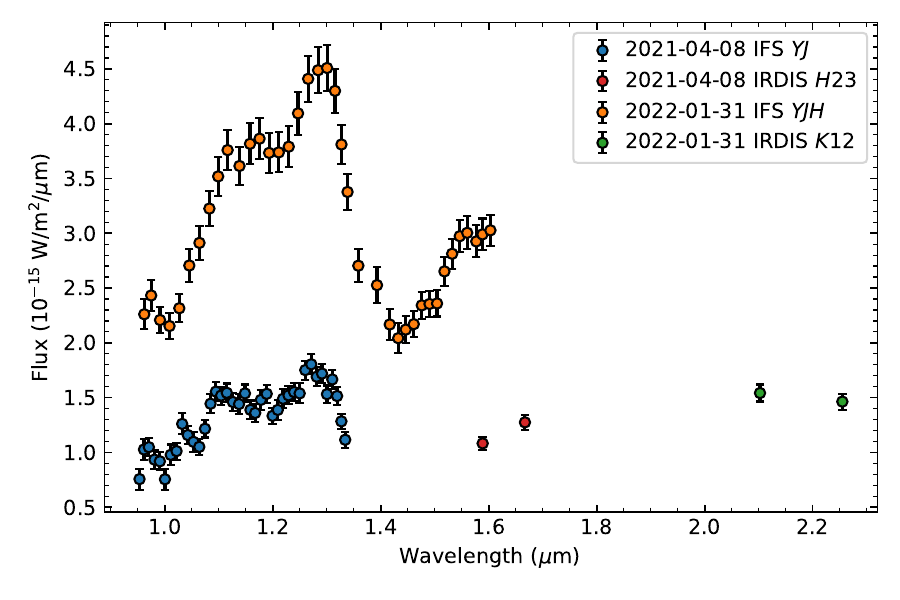}
    \caption{} \label{fig:hd112863noDTTSnoCC}
    \end{subfigure}
    \begin{subfigure}[t]{0.49\linewidth}
    \includegraphics[width=\textwidth]{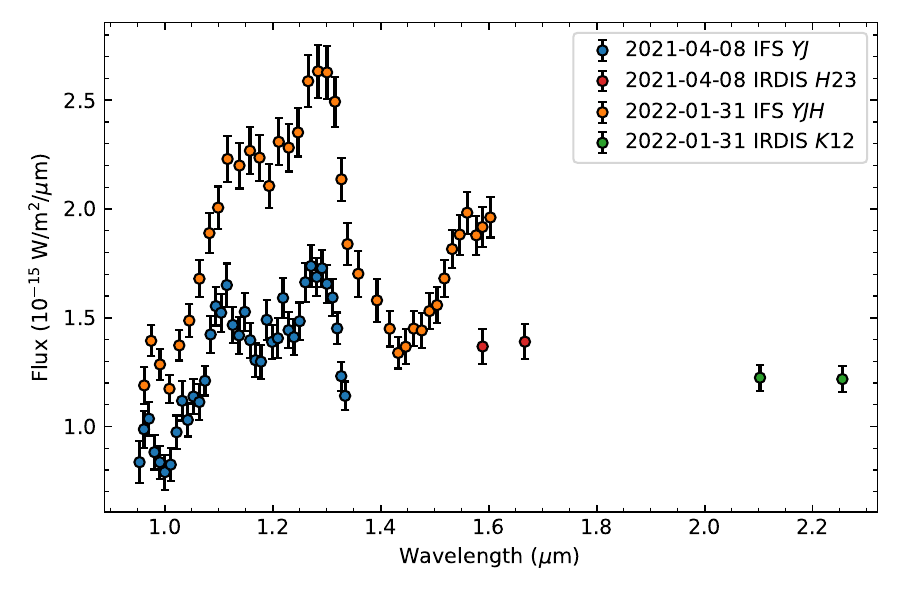}
    \caption{} \label{fig:hd112863wDTTSnoCC}
    \end{subfigure}\par
    \begin{subfigure}[t]{0.49\linewidth}
    \includegraphics[width=\textwidth]{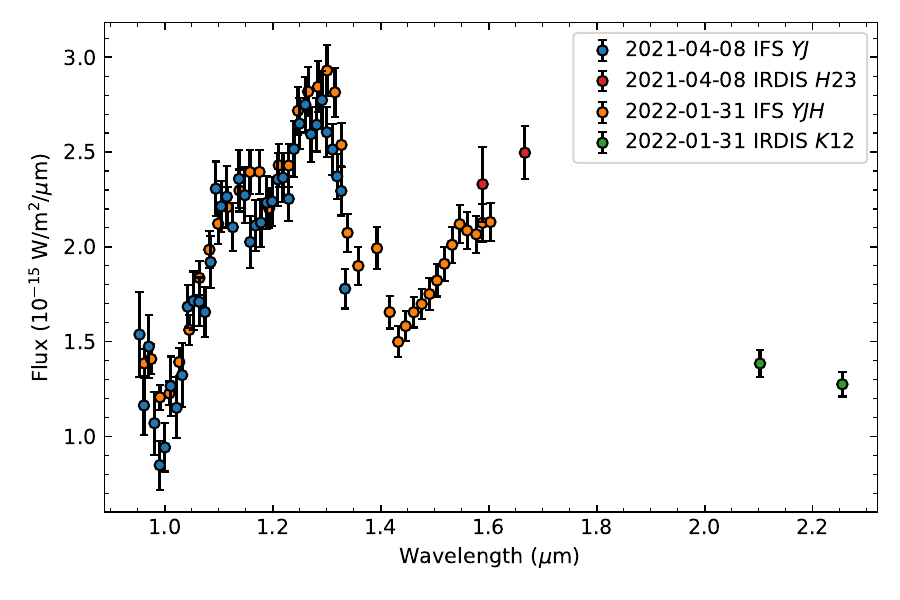}
    \caption{} \label{fig:hd112863DTTSandCC}
    \end{subfigure}
    \caption{\textbf{(a)} The original flux spectra of HD~112863~B, before accounting for the effects of weather and the coronagraphic attenuation of the companion.  \textbf{(b)} The partially corrected flux spectra of HD~112863~B, after accounting for weather effects (see Sect.~\ref{sec:observs} and Sect.~\ref{subsubsec:weather}). \textbf{(c)} The final flux spectra of HD~112863~B, after accounting for both the effects of weather and the coronagraphic attenuation of the companion (see Sect.~\ref{subsubsec:coron}).}
    \label{fig:hd112863_all_spectra_versions}
\end{figure*}

\begin{figure*}[!h]
    \centering
    \begin{subfigure}[t]{0.49\linewidth}
    \includegraphics[width=\textwidth]{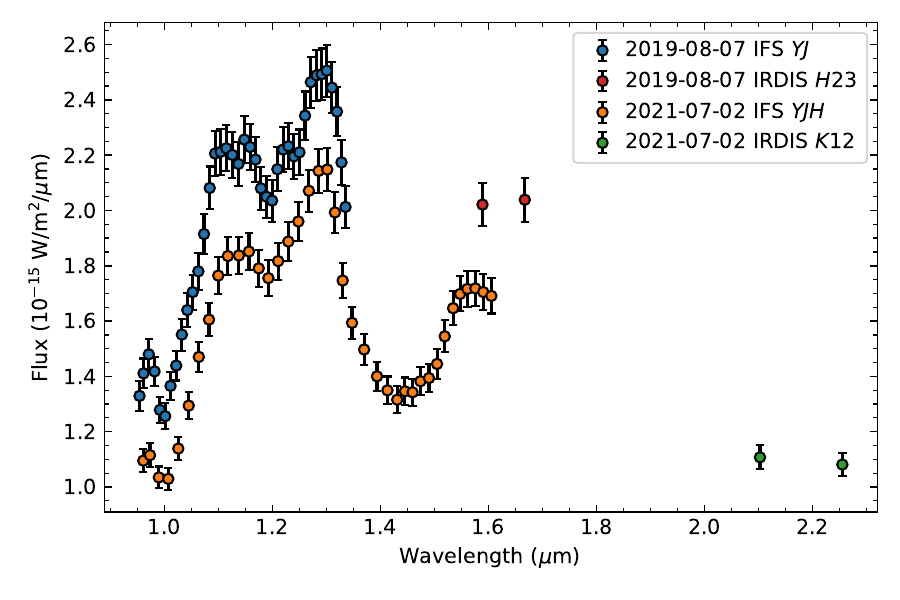} 
    \caption{} \label{fig:hd206505noDTTS}
    \end{subfigure}
    \begin{subfigure}[t]{0.49\linewidth}
    \includegraphics[width=\textwidth]{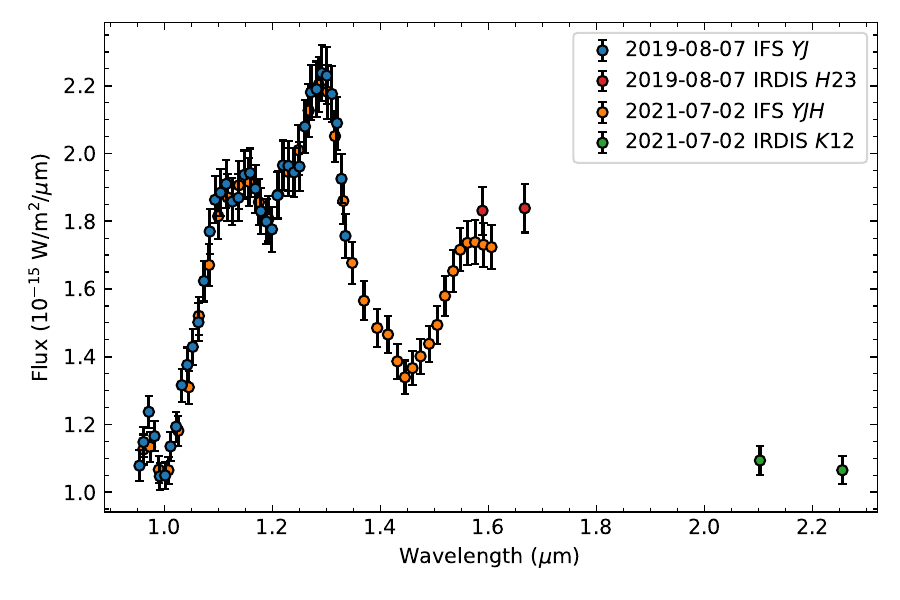}
    \caption{} \label{fig:hd206505DTTS}
    \end{subfigure}
    \caption{The flux spectra of HD~206505~B, before \textbf{(a)} and after \textbf{(b)} accounting for the effects of weather on the datasets (see Sect.~\ref{sec:observs} and Sect.~\ref{subsubsec:weather}).} \label{fig:hd206505_all_spectra_versions}
\end{figure*}

Shown here are the status of the spectra of HD~112863~B and HD~206505~B at each step of the data reduction process, as detailed in Sect.~\ref{sec:datareduc}.  The spectra of both companions required correction for the weather during observations (see Sect.~\ref{subsubsec:weather} and Appendix~\ref{appendix_dtts}), the results of which are shown in Fig.~\ref{fig:hd112863wDTTSnoCC} and Fig.~\ref{fig:hd206505DTTS}.  Additionally, the spectrum of HD~112863~B required correction for the transmittance of the SPHERE coronagraph (see Sect.~\ref{subsubsec:coron} and Appendix~\ref{appendix_corons}), the result of which is shown in Fig.~\ref{fig:hd112863DTTSandCC}.

\FloatBarrier

\section{DTTS data} \label{appendix_dtts}

\begin{figure*}[!h]
    \centering
    \includegraphics[width=0.49\linewidth]{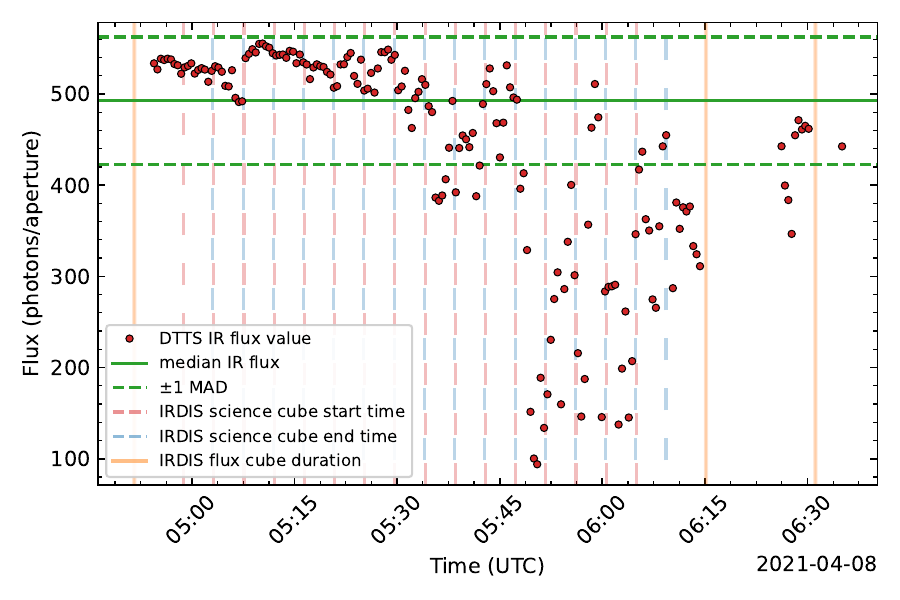}
    \includegraphics[width=0.49\linewidth]{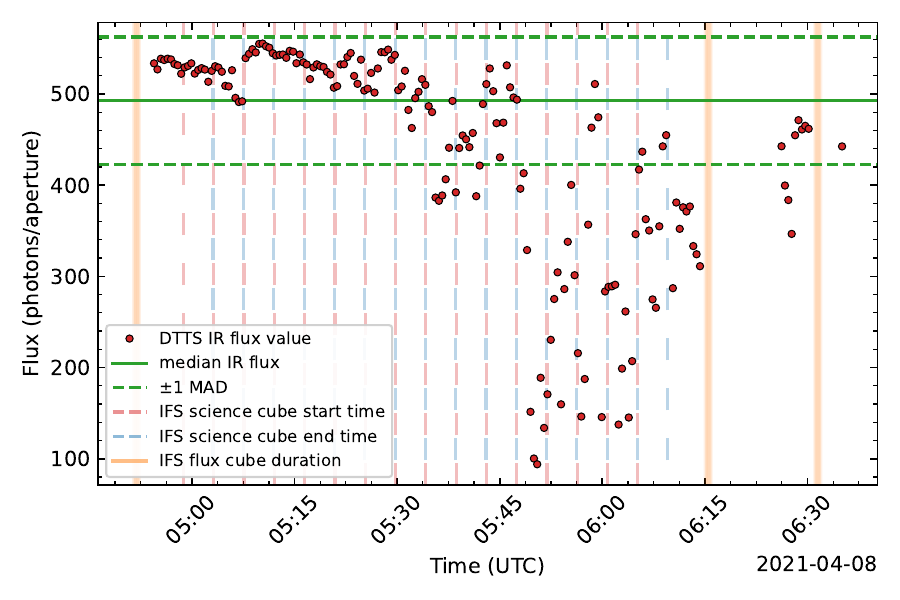}
    \caption{The DTTS flux measurements of HD~112863 during the 2021-04-07 IRDIS $H23$ (\textbf{left}) and IFS $YJ$ (\textbf{right}) observations.  \textit{Green lines} indicate the median and $\pm1$ MAD of the measurements, which are the criteria used when excluding science frames based on weather conditions (see Sect.~\ref{subsubsec:weather}).}
    \label{fig:hd112863_H23_YJ_DTTS}
\end{figure*}

\begin{figure*}[!h]
    \centering
    \includegraphics[width=0.49\linewidth]{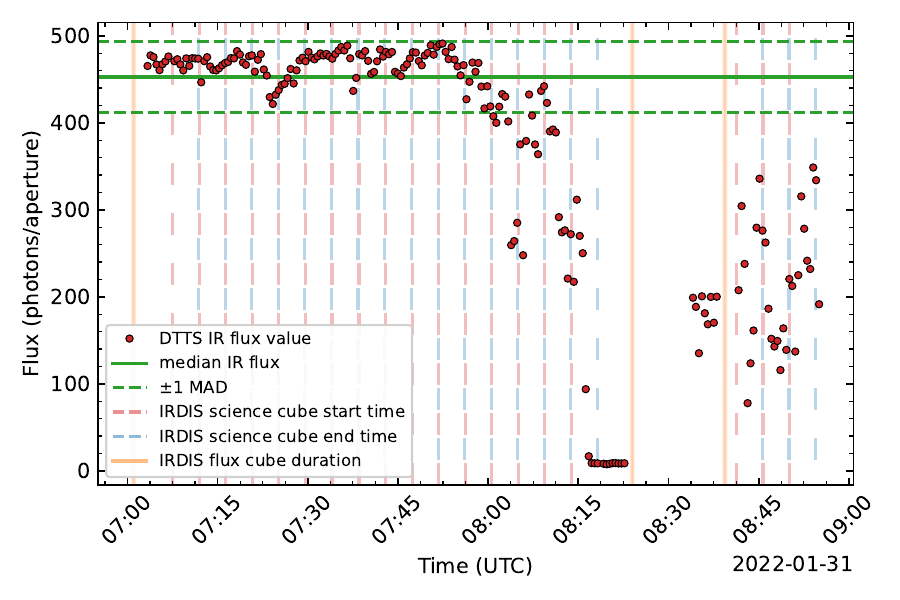}
    \includegraphics[width=0.49\linewidth]{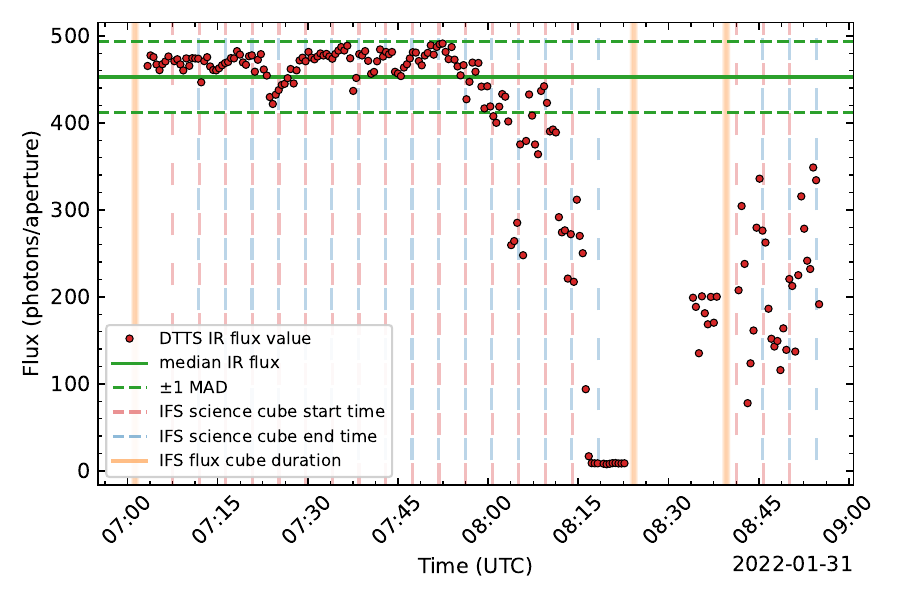}
    \caption{Same as Fig.~\ref{fig:hd112863_H23_YJ_DTTS} but for HD~112863 during the 2022-01-30 IRDIS $K12$ (\textbf{left}) and IFS $YJH$ (\textbf{right}) observations.}
    \label{fig:hd112863_K12_YJH_DTTS}
\end{figure*}

\begin{figure*}[!h]
    \centering
    \includegraphics[width=0.49\linewidth]{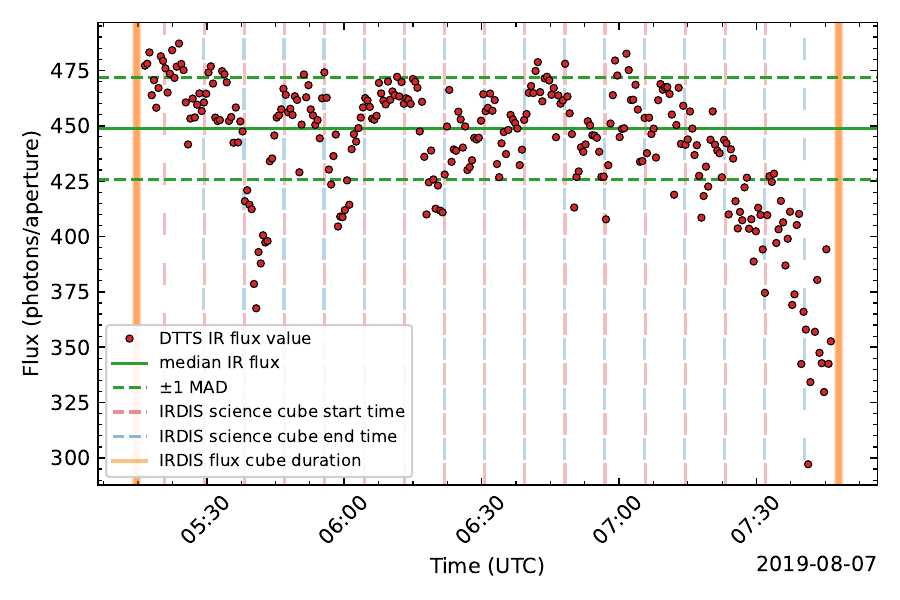}
    \includegraphics[width=0.49\linewidth]{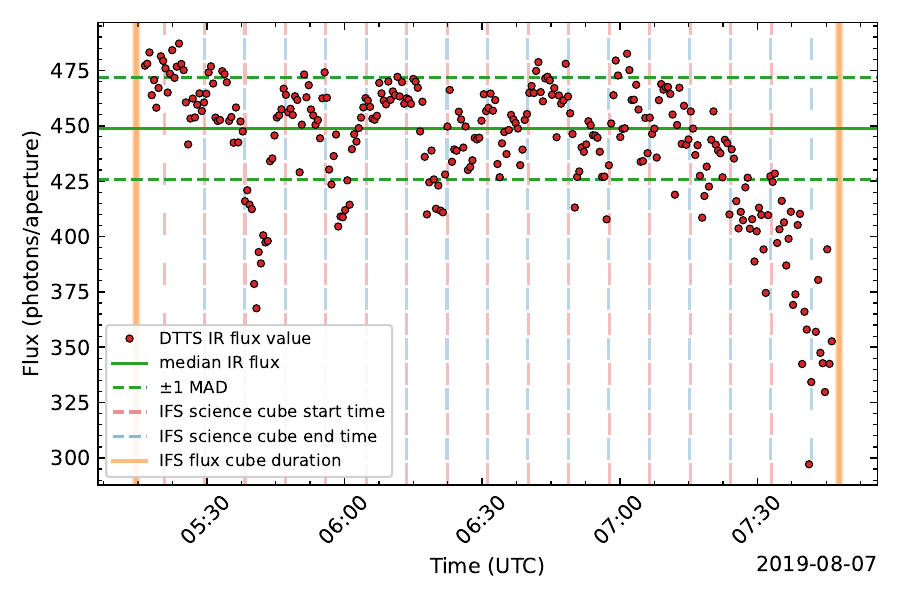}
    \caption{Same as Fig.~\ref{fig:hd112863_H23_YJ_DTTS} but for HD~206505 during the 2019-01-31 IRDIS $H23$ (\textbf{left}) and IFS $YJ$ (\textbf{right}) observations.}
    \label{fig:hd206505_H23_YJ_DTTS}
\end{figure*}

\begin{figure*}[!h]
    \centering
    \includegraphics[width=0.49\linewidth]{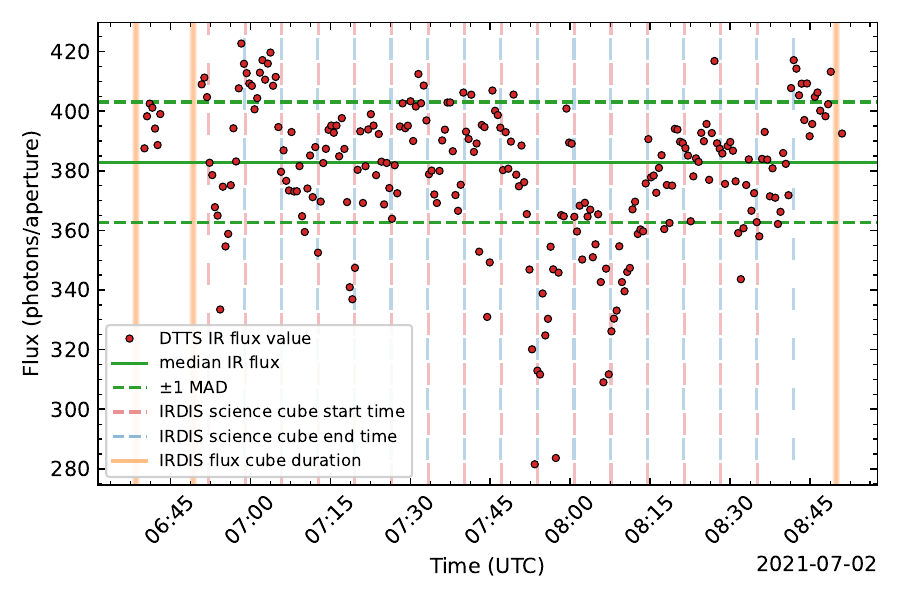}
    \includegraphics[width=0.49\linewidth]{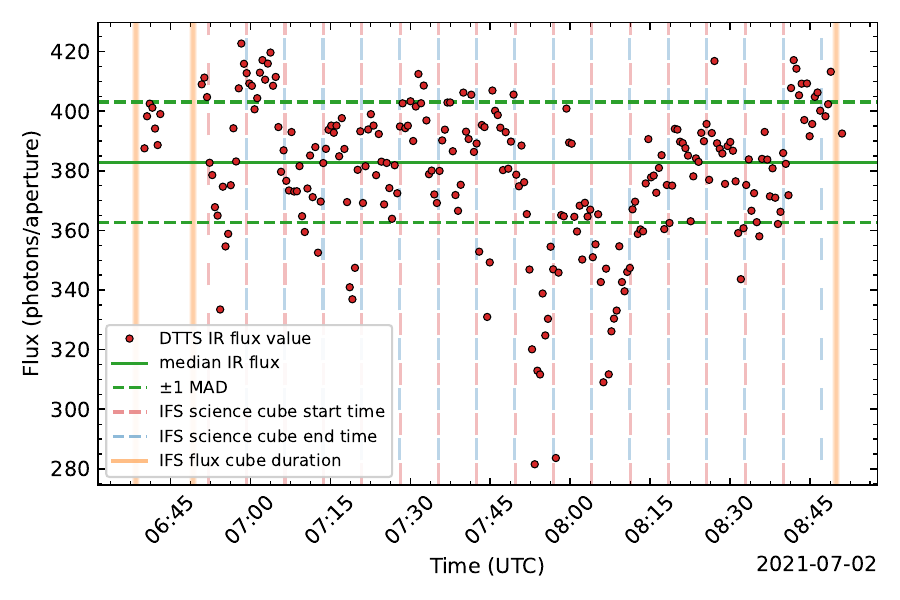}
    \caption{Same as Fig.~\ref{fig:hd112863_H23_YJ_DTTS} but for HD~206505 during the 2021-07-01 IRDIS $K12$ (\textbf{left}) and IFS $YJH$ (\textbf{right}) observations.}
    \label{fig:hd206505_K12_YJH_DTTS}
\end{figure*}

Included here are the flux measurements recorded by the SPARTA computer on the DTTS of SPHERE, for both epochs of HD~112863 and HD~206505 observations.  The DTTS flux measurements for the two epochs of HD~112863 observations are shown in Fig.~\ref{fig:hd112863_H23_YJ_DTTS} and Fig.~\ref{fig:hd112863_K12_YJH_DTTS}, while the DTTS flux measurements for the two epochs of HD~206505 observations are shown in Fig.~\ref{fig:hd206505_H23_YJ_DTTS} and Fig.~\ref{fig:hd206505_K12_YJH_DTTS}.  The durations of the science and flux cubes are also shown for each set of observations.

\FloatBarrier

\section{Coronagraph transmission profiles} \label{appendix_corons}

\begin{figure*}[!h]
    \centering
    \includegraphics[width=0.49\linewidth]{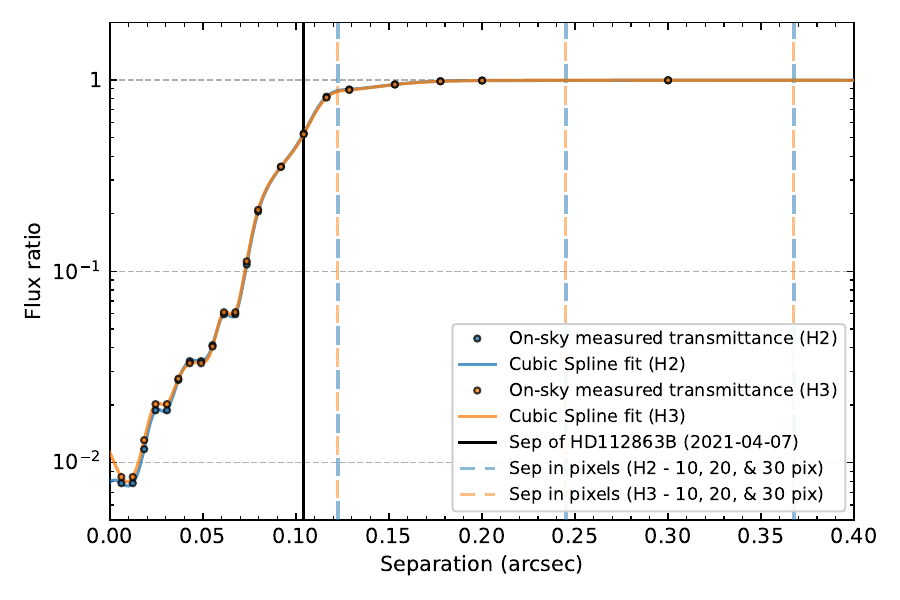}
    \includegraphics[width=0.49\linewidth]{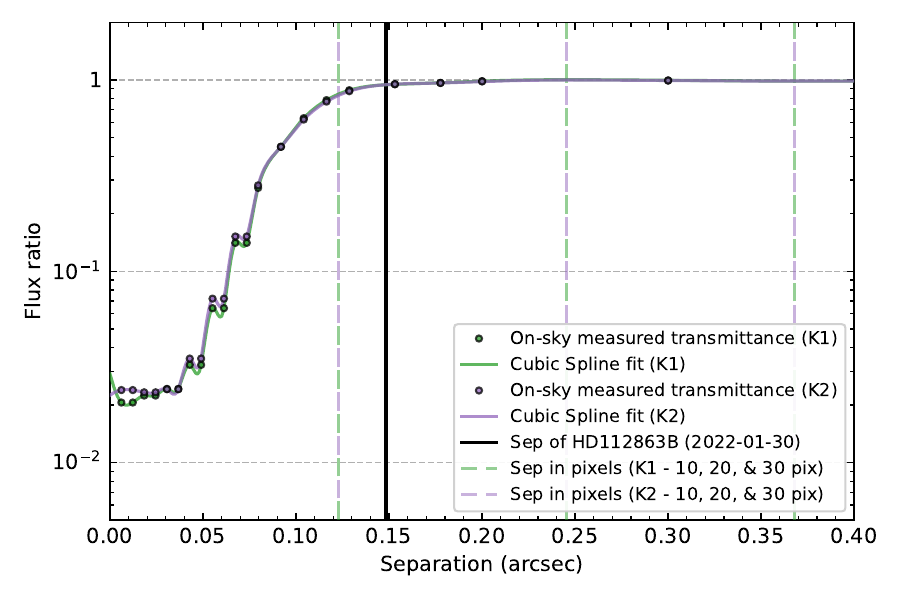}
    \caption{The coronagraph transmission profiles for IRDIS $H23$ (\textbf{left}) and $K12$ (\textbf{right}).  Shown here are the measured values taken on-sky with seeing conditions of $\sim0.8$~arcsec (A. Vigan), as well as the cubic spline interpolation to these values.  These interpolations were treated as radial profiles to create coronagraphic transmission images, in order to directly correct the HD~112863 science frames (see Sect.~\ref{subsubsec:coron}).  Also shown are the equivalent $\rho$ at $10$~pixels, $20$~pixels, $30$~pixels, accounting for the difference in pixel scale between each band.}
    \label{fig:hd112863_coron_IRDIS}
\end{figure*}

\begin{figure}[!h]
    \centering
    \includegraphics[width=0.5\textwidth]{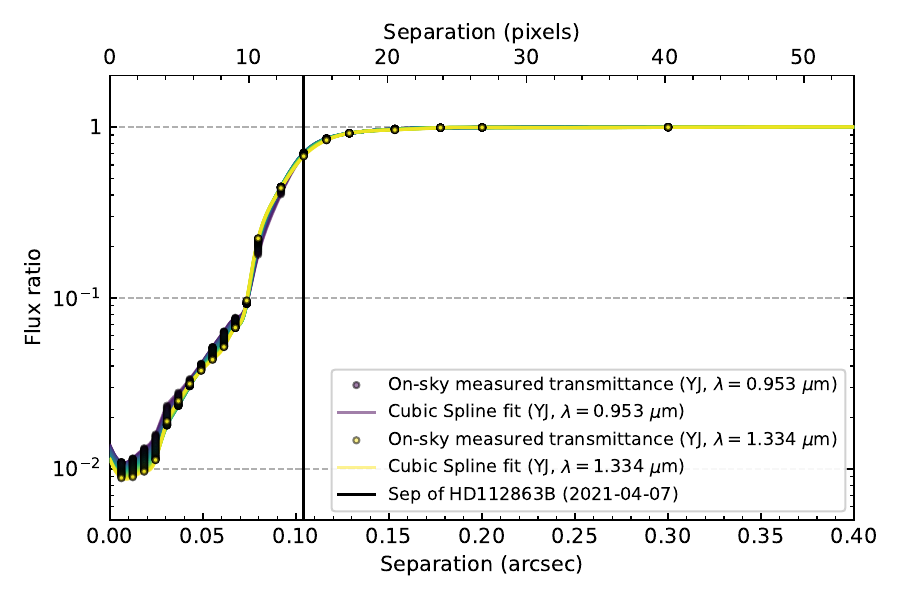}
    \caption{Same as Fig.~\ref{fig:hd112863_coron_IRDIS} but for the coronagraph transmission profiles for IFS $YJ$. 
 The measurements and interpolations for the first and last wavelength of IFS $YJ$ are highlighted.}
    \label{fig:hd112863_coron_YJ}
\end{figure}

\begin{figure}[!h]
    \centering
    \includegraphics[width=0.5\textwidth]{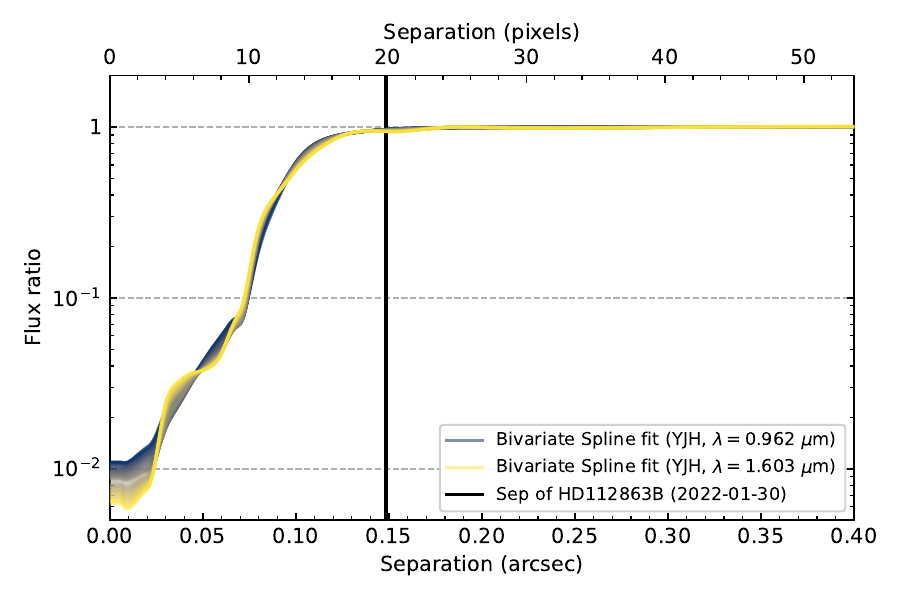}
    \caption{The coronagraph transmission profiles for IFS $YJH$.  Shown here are the results of the bivariate spline interpolation to the IFS $YJ$ and IRDIS $H23$ measured values, at each of the IFS $YJH$ wavelengths.  The interpolations for the first and last wavelength of IFS $YJH$ are highlighted. The interpolations were treated as radial profiles to create coronagraphic transmission images, in order to directly correct the HD~112863 science frames (see Sect.~\ref{subsubsec:coron}).}
    \label{fig:hd112863_coron_YJH}
\end{figure}

Included here are the on-sky-measured transmission profiles of the N\_ALC\_YJH\_S coronagraph integrated in SPHERE.  The measured transmission values, along with the interpolated profiles, are shown in Fig.~\ref{fig:hd112863_coron_IRDIS} for IRDIS $H23$ and $K12$ and in Fig.~\ref{fig:hd112863_coron_YJ} for IFS $YJ$.  Because transmission measurements of the coronagraph over the IFS $YJH$ wavelength range are not available, we interpolated between the measured transmission values for IRDIS $H23$ and IFS $YJ$ to generate coronagraph transmission profiles for IFS $YJH$, shown in Fig.~\ref{fig:hd112863_coron_YJH}.

\FloatBarrier

\section{Updated relative astrometry and photometry} \label{appendix_relastrom}

We recomputed the relative astrometry and photometry of HD~112863~B and HD~206505~B, with the results shown in Table~\ref{tab:updtd_astro_photo}; the data reduction approach used in \cite{Rickman2024} did not include some of the steps described in Sect.~\ref{subsubsec:weather} and Sect.~\ref{subsubsec:coron}, resulting in a flux offset between the two epochs of spectrophotometric observations for both companions (see Appendix~\ref{appendix_uncorrspecs}).  We note that the updated relative astrometry is consistent with the values presented in \cite{Rickman2024} within $\pm1\sigma$.

\begin{table*}[!h]
    \centering
        \caption{Updated relative astrometry and photometry of HD~112863~B and HD~206505~B.}   
    \begin{tabular}{ccccccc}
    \hline
    \hline
        Companion & Date (yyyy-mm-dd) & Filter & $\rho$ (mas) & $\theta$ (deg) & Contrast & Abs. Mag \\
    \hline
        HD~112863~B & 2021-04-07 & H2 & $104.2\pm5.7$ & $309.6\pm2.1$ & $7.54\pm0.17$ & $11.48\pm0.09$ \\
        HD~112863~B & 2021-04-07 & H3 & $104.1\pm6.9$ & $310.1\pm2.4$ & $7.35\pm0.25$ & $11.22\pm0.06$ \\
        HD~112863~B & 2022-01-30 & K1 & $147.3\pm3.6$ & $328.8\pm1.2$ & $6.99\pm0.05$ & $10.97\pm0.05$ \\
        HD~112863~B & 2022-01-30 & K2 & $149.4\pm5.1$ & $327.3\pm1.3$ & $6.75\pm0.10$ & $10.77\pm0.05$ \\
        HD~206505~B & 2019-08-06 & H2 & $374.9\pm3.1$ & $277.5\pm0.5$ & $7.662\pm0.003$ & $11.38\pm0.04$ \\ 
        HD~206505~B & 2019-08-06 & H3 & $376.3\pm3.1$ & $276.2\pm0.5$ & $7.533\pm0.003$ & $11.19\pm0.04$ \\ 
        HD~206505~B & 2021-07-01 & K1 & $404.3\pm3.1$ & $274.3\pm0.5$ & $7.196\pm0.003$ & $10.86\pm0.04$ \\ 
        HD~206505~B & 2021-07-01 & K2 & $403.1\pm3.1$ & $274.3\pm0.5$ & $6.954\pm0.003$ & $10.60\pm0.04$ \\ 
        \hline
    \end{tabular}
    \tablefoot{These values replace those presented in \cite{Rickman2024}, since they are from the updated data reduction approach presented in this paper, that includes both weather correction (see Sect.~\ref{subsubsec:weather}) and coronagraph transmission correction (see Sect.~\ref{subsubsec:coron}).  Otherwise, note that these photometric values were obtained following the same procedure as described in Sect.~$3.3$ of \cite{Rickman2024}, and are not the values extracted by TRAP (as shown in Fig.~\ref{fig:hd112863DTTSandCC} and Fig.~\ref{fig:hd206505DTTS}).}\label{tab:updtd_astro_photo}
\end{table*}

\FloatBarrier

\section{Updated color-magnitude diagram} \label{appendix_CMD}

Since the data reduction approach used in \cite{Rickman2024} did not include some of the steps described in Sect.~\ref{subsubsec:weather} and Sect.~\ref{subsubsec:coron}, we generated a new CMD, shown in Fig.~\ref{fig:cmd}, that uses the updated photometry of HD~112863~B and HD~206505~B from this work (see Appendix~\ref{appendix_relastrom}, specifically Table~\ref{tab:updtd_astro_photo}).

\begin{figure}[!h]
    \centering
    \includegraphics[width=0.5\textwidth]{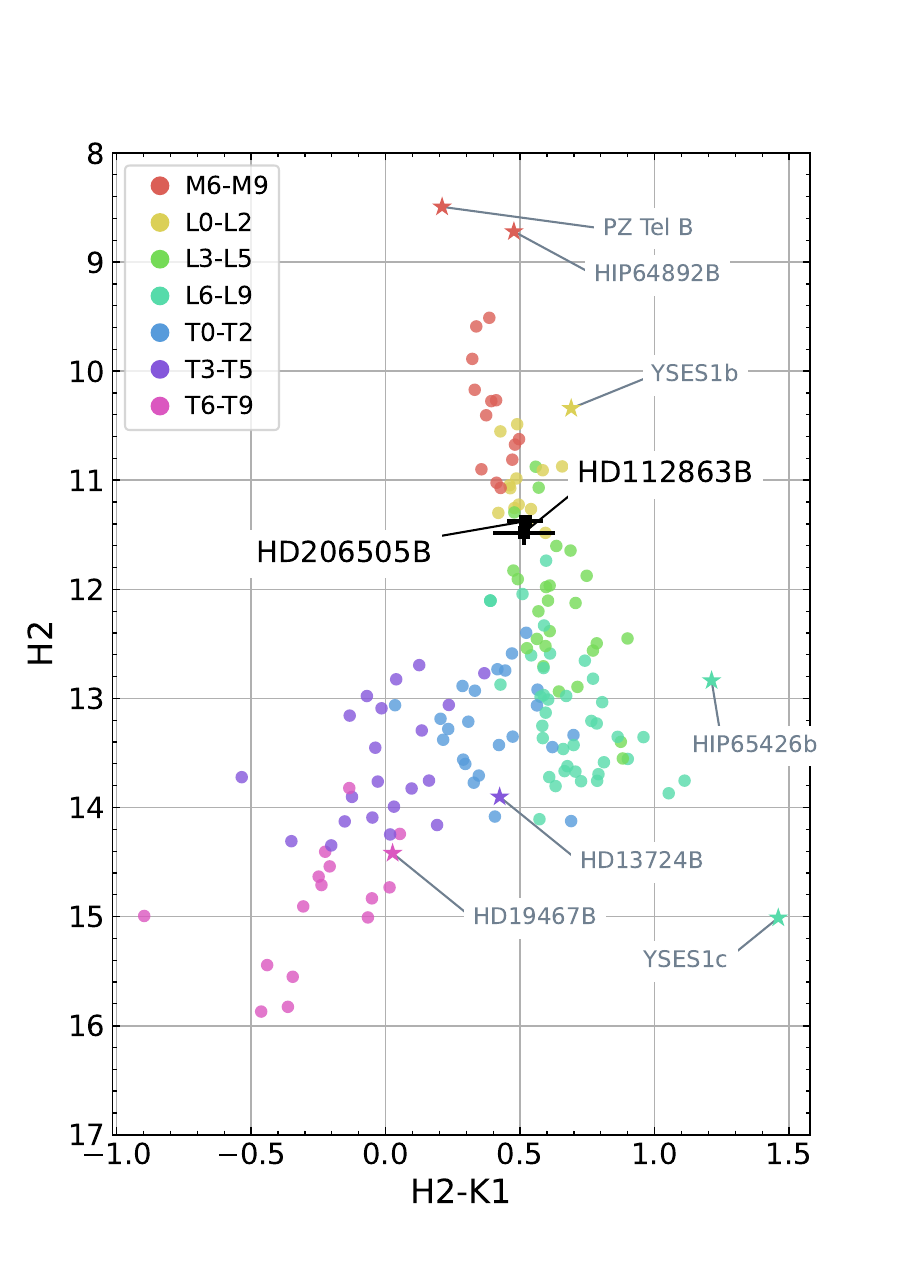}
    \caption{The updated CMD (compared to Fig.~4 of \citealp{Rickman2024}) showing HD~112863~B and HD~206505~B (\textit{black squares}, from Table~\ref{tab:updtd_astro_photo}) in comparison to the population of field brown dwarfs (\textit{circle symbols}), as well as some notable substellar companions (\textit{star symbols}). Field brown dwarfs are color-coded by spectral classification.}
    \label{fig:cmd}
\end{figure}

\FloatBarrier

\section{Updated orbit fits} \label{appendix_orbitfits}

Using the updated relative astrometry shown in Table~\ref{tab:updtd_astro_photo}, we obtain new orbit fits of both systems using the orbit-fitting code \texttt{orvara} \citep{2021AJ....162..186B}. We use the RV measurements, along with the HIPPARCOS-\textit{Gaia} catalog of accelerations \citep[HGCA,][]{2021ApJS..254...42B} as outlined in \cite{Rickman2022} and \cite{Rickman2024}, and incorporate the updated astrometry as presented in this paper. 

For both systems, we performed an orbit fit using a parallel-tempered MCMC with 15~temperatures. For each temperature, we used 100~walkers with 40000~steps per walker thinned by a factor of 50. We used a log-flat prior on the host star mass in order to derive the mass of the star dynamically. The resulting updated dynamical masses and orbital parameters are all in agreement with \cite{Rickman2024}, and are shown in Table~\ref{tab:orbitparameters}. The dynamical mass of HD~206505~B did not change from \cite{Rickman2024}, and the dynamical mass of HD~112863~B changed fractionally with the previous value of $77.1^{+2.9}_{-2.8}~M_{\rm{Jup}}$ to now $77.7^{+3.4}_{-3.3}~M_{\rm{Jup}}$, which is due to HD~112863~B being within the IWA of the coronagraph and applying the coronagraphic transmission factor as described in Sect.~\ref{subsubsec:coron}. Despite a slight increase in the uncertainty folded in due to this transmission correction, the orbital parameters are well-defined due to the extensive orbital phase coverage of the RVs \citep{Rickman2024}, and are all in agreement with the previously derived values.

\FloatBarrier

\begin{table*}[h!]
    \caption{MCMC orbital posteriors for the orbital fits of each system using \texttt{orvara} \citep{2021AJ....162..186B}.}
    \centering
    \begin{tabular}{ccccc}
         \hline \hline 
         Parameters & Units & Prior & HD~112863 & HD~206505 \\
         \hline
         & & Fitted Parameters \\
         \hline
         Companion mass $M_{\rm{comp}}$ & $M_{\rm{Jup}}$ & $1/M_{\rm{comp}}$ (log-flat) & ${77.7}_{-3.3}^{+3.4}$ & $79.8\pm1.8$ \\
         Host-star mass $M_{\rm{host}}$ & $M_{\rm{\odot}}$ & $1/M_{\rm{host}}$ (log-flat) & ${0.90}_{-0.05}^{+0.06}$ & $0.97\pm0.03$ \\
         Parallax $\varpi$ & mas & $1/\varpi$ (log-flat) & ${26.955}_{-0.003}^{+0.002}$ & $22.768\pm0.003$ \\
         Inclination $i$ & $\degree$ & $\sin i$ & ${60.55}_{-0.64}^{+0.65}$ & ${110.21}_{-0.75}^{+0.73}$ \\
         Semimajor axis $a$ & AU & $1/a$ (log-flat) & ${7.69}_{-0.15}^{+0.16}$ & ${13.92}_{-0.16}^{+0.17}$ \\
         Jitter $\sigma$ & ms$^{-1}$ & $1/\sigma$ (log-flat) & ${15.3}_{-1.2}^{+1.4}$ & ${4.1}_{-1.2}^{+1.1}$ \\
         $\sqrt{e}\sin\omega$ & & $\mathcal{U}(-1,1)$ & ${-0.546}_{-0.004}^{+0.005}$ & $0.512\pm0.009$ \\
         $\sqrt{e}\cos\omega$ & & $\mathcal{U}(-1,1)$ & $-0.219\pm0.008$ & $0.572\pm0.007$ \\
         PA of the ascending node $\Omega$ & $\degree$ & $\mathcal{U}(-180,180)$ & ${346.29}_{-0.91}^{+0.88}$ & ${97.57}_{-0.38}^{+0.37}$ \\
         \hline
         & & Derived Parameters \\
         \hline 
         Orbital Period $P$ & years & & $21.60\pm0.05$ & ${50.9}_{-1.5}^{+1.7}$ \\
         Eccentricity $e$ & & & $0.346\pm0.004$ & ${0.597}_{-0.008}^{+0.009}$ \\
         Argument of periastron $\omega$ & $\degree$ & & ${248.13}_{-0.85}^{+0.82}$ & ${42.26}_{-0.73}^{+0.72}$ \\
         Time of Periastron $T_0$ & JD & & $2458702\pm12$ & ${2473383}_{-558}^{+633}$ \\
         Semimajor axis & mas & & ${207.4}_{-4.2}^{+4.3}$ & ${317.0}_{-3.6}^{+4.0}$ \\
         Mass ratio $q$ & $M_{\rm{comp}}/M_{\rm{host}}$ & & $0.082\pm0.002$ & $0.079\pm0.001$ \\
         \hline \\
    \end{tabular}
    \label{tab:orbitparameters}
\end{table*}

\begin{figure*}[h!]
    \centering
    \includegraphics[width=0.48\linewidth]{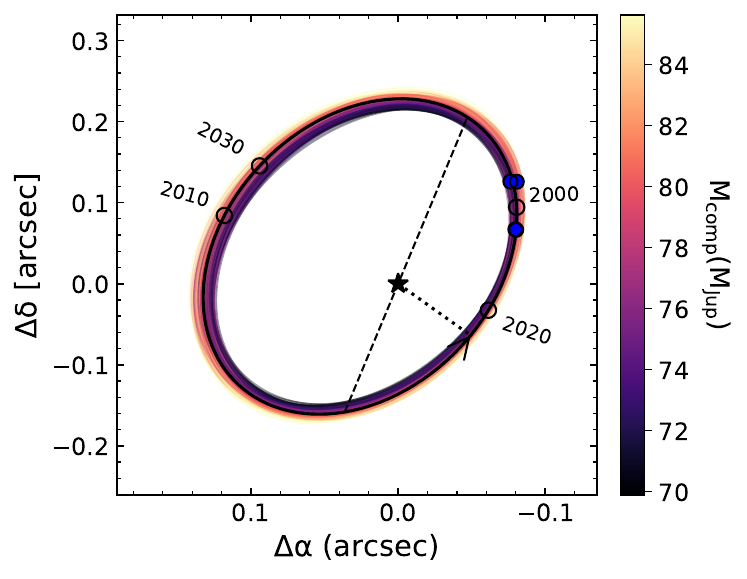}
    \includegraphics[width=0.49\linewidth]{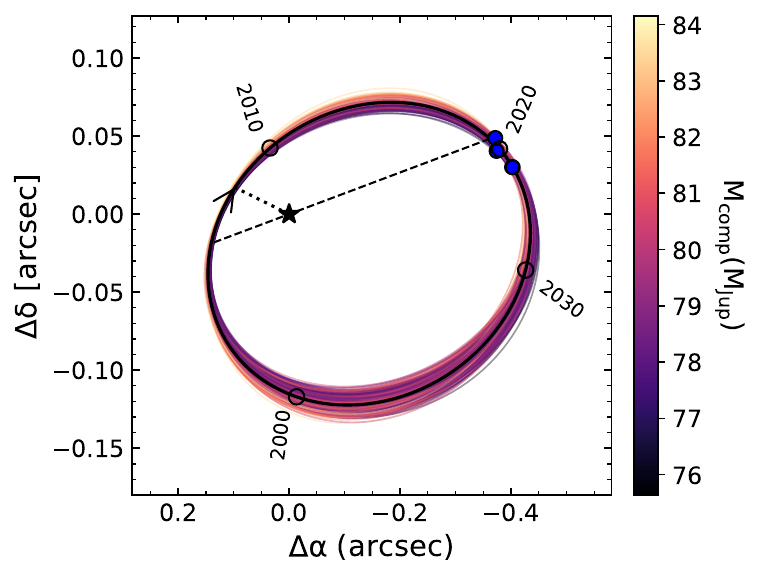}
    \caption{Updated versions of the top right panels of Fig.~2 and Fig.~3 from \cite{Rickman2024}; the measured astrometric positions (\textit{blue data points}) and orbits of HD~112863~B (\textbf{left}) and HD~206505~B (\textbf{right}) relative to their host stars. The corresponding masses from the orbit fits are represented by the color bar to the right of each plot.}
    \label{fig:orbits}
\end{figure*}

\begin{figure*}[h!]
    \centering  
    \includegraphics[width=\linewidth]{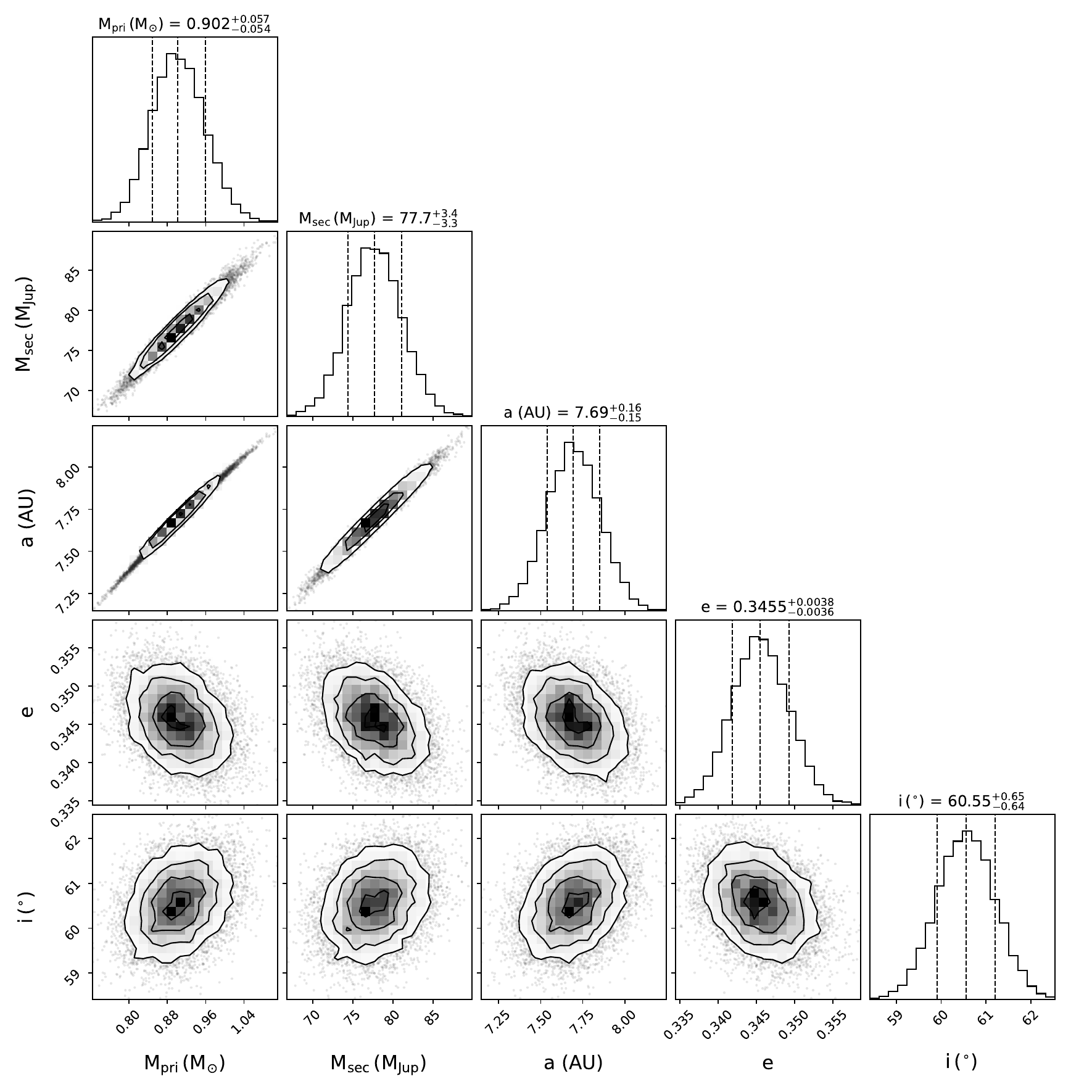}
    \caption{Corner plot of orbital parameters of HD~112863~B (updated version of Fig.~B.1 in \citealp{Rickman2024}).}
    \label{fig:HD112863-cornerplot}
\end{figure*}

\begin{figure*}[h!]
    \centering 
    \includegraphics[width=\linewidth]{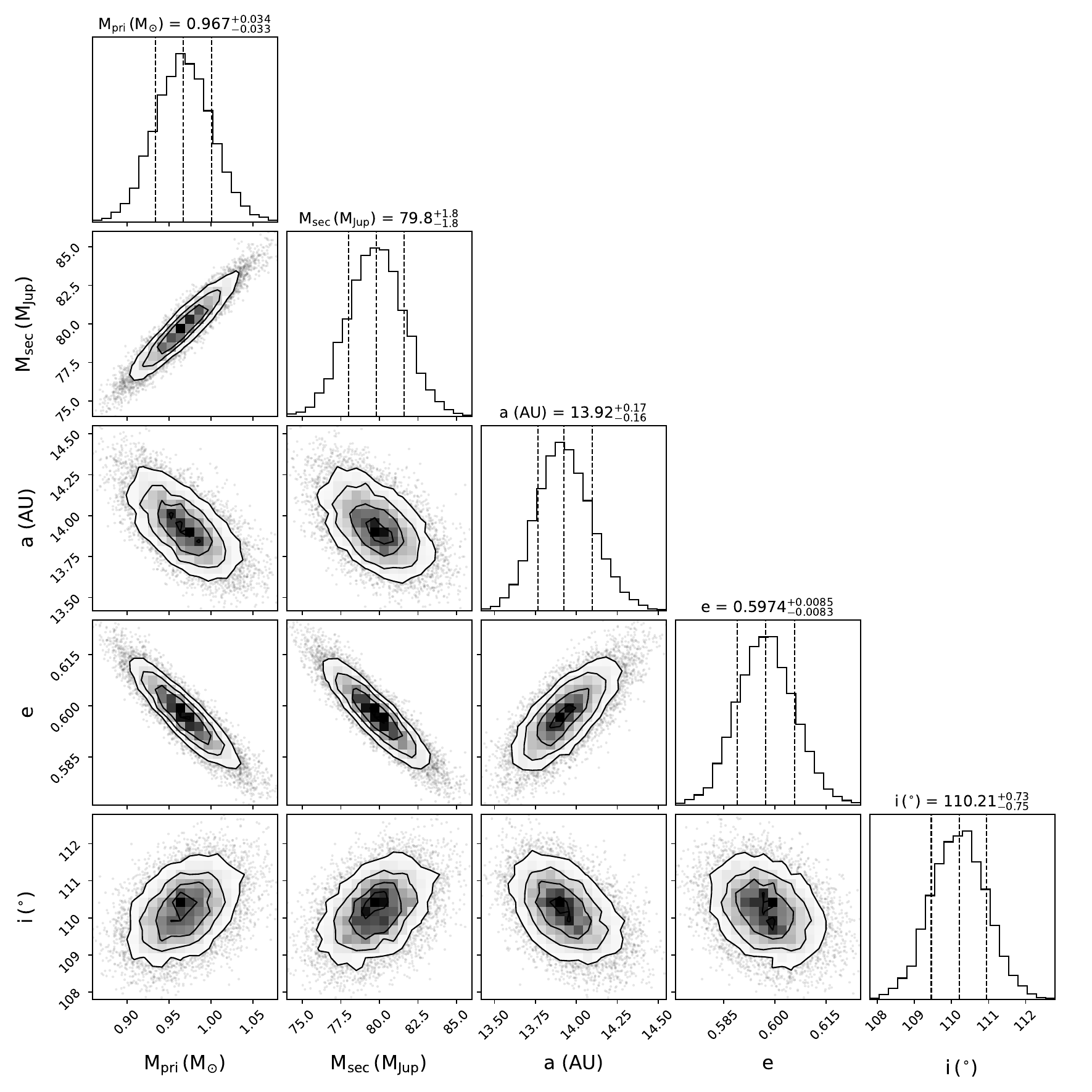}
    \caption{Corner plot of orbital parameters of HD~206505~B (updated version of Fig.~B.2 in \citealp{Rickman2024}).}
    \label{fig:HD206505-cornerplot}
\end{figure*}

\FloatBarrier

\section{Atmospheric parameters posterior distributions} \label{appendix_atmoposts}
Shown here are the posterior distributions of the parameters from the atmospheric model fits to HD~112863~B and HD~206505~B, using both the BT-Settl and Sonora Diamondback model grids (see Sect.~\ref{subsec:atmofits}).  

\begin{figure*}[!h]
    \centering
    \includegraphics[width=0.46\linewidth]{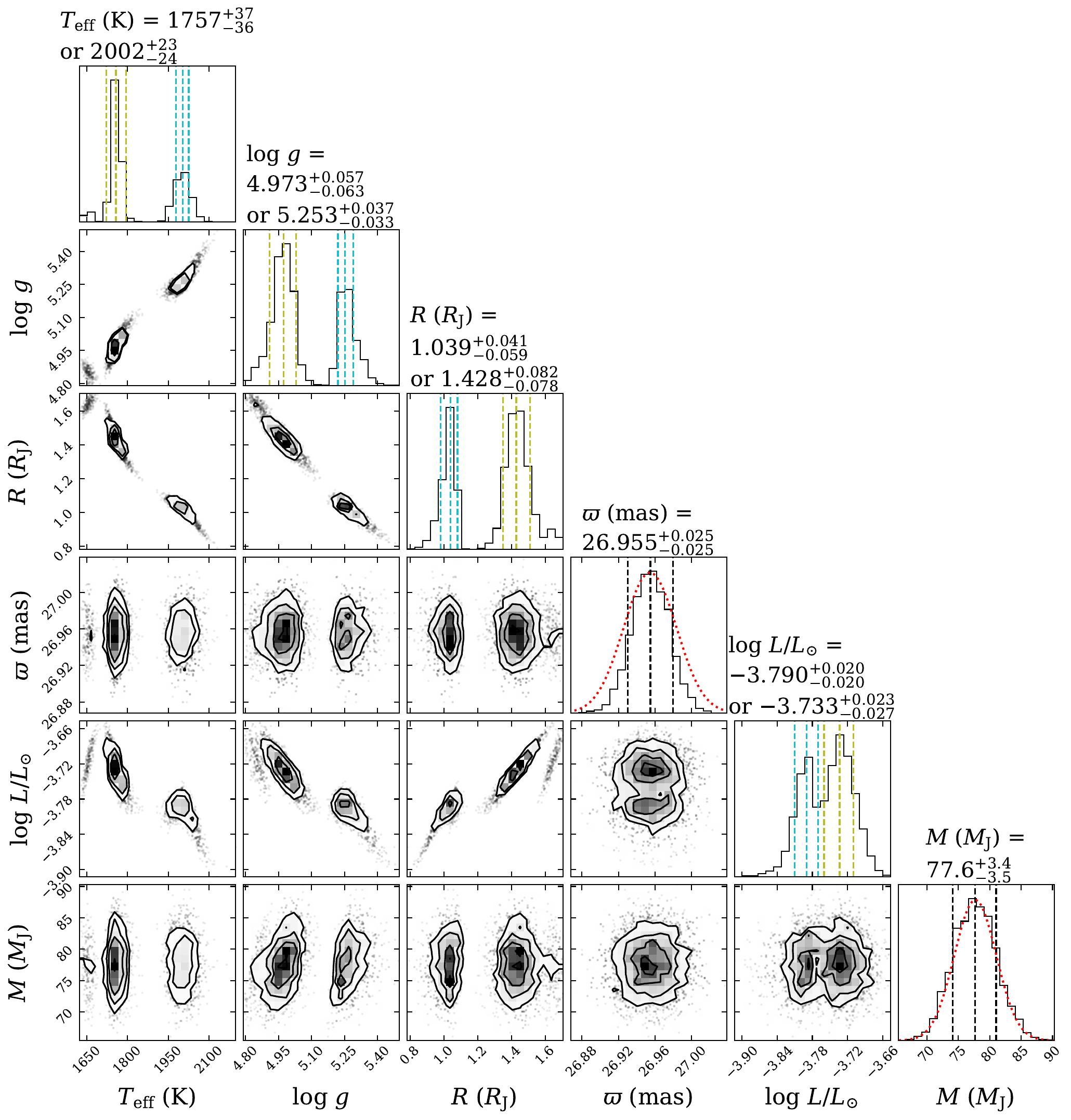}
    \caption{The posterior distributions of the physical and derived parameters from the BT-Settl atmospheric model fit to HD~112863~B (see Fig.~\ref{fig:hd112863_btsettl_main}).  Included are the median values and $1\sigma$ credible intervals for both the first mode (\textit{dashed olive lines}) and the second mode (\textit{dashed cyan lines}) of the bimodal posteriors.  The priors used for $\varpi$ and $M$ are also included (\textit{red dotted lines}).}
    \label{fig:hd112863_btsettl_posts}
\end{figure*}

\begin{figure*}[!h]
    \centering
    \includegraphics[width=0.56\linewidth]{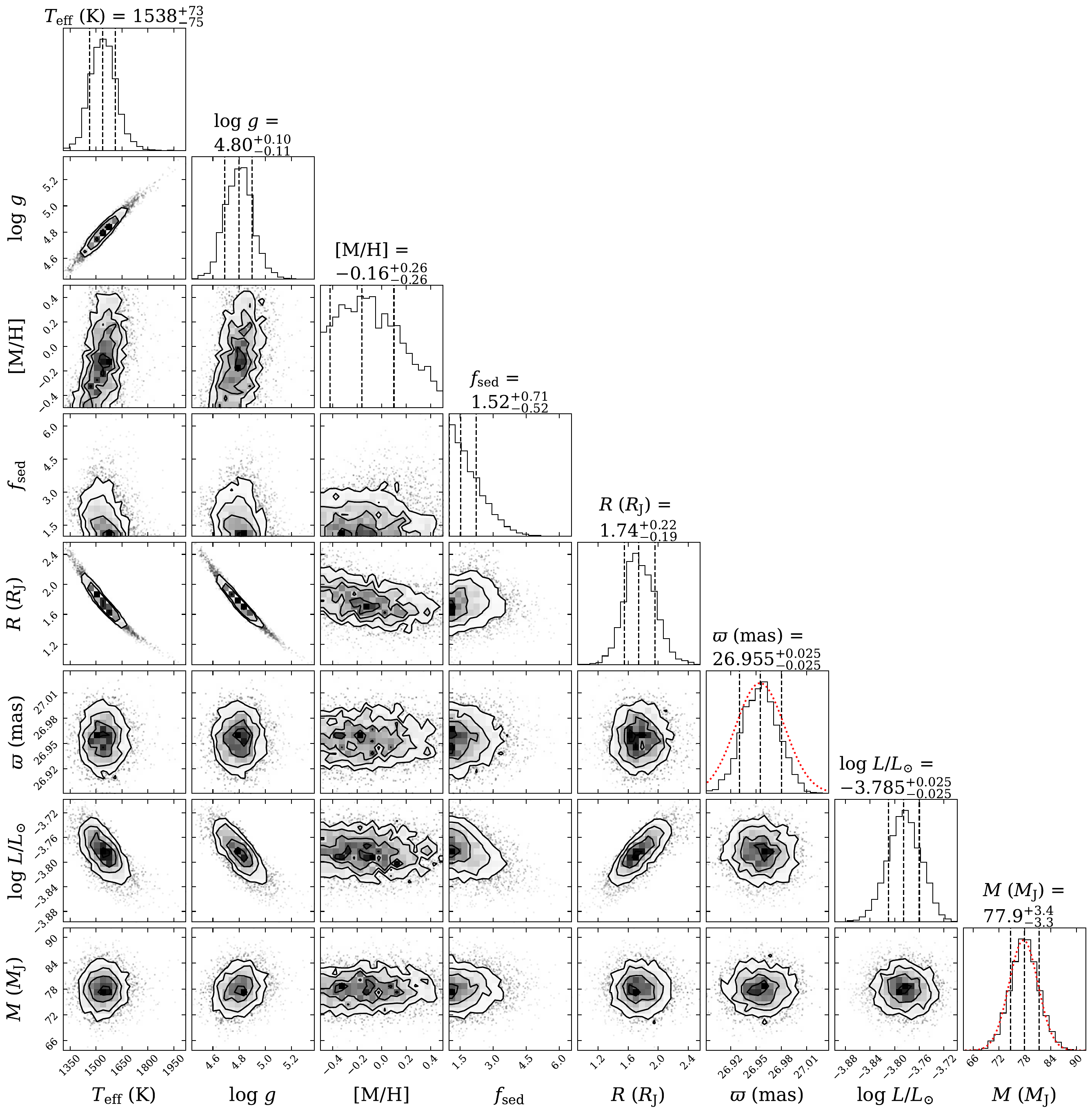}
    \caption{The posterior distributions of the physical and derived parameters from the Sonora Diamondback atmospheric model fit to HD~112863~B (see Fig.~\ref{fig:hd112863_sonoradiamond_main}).}
    \label{fig:hd112863_sonoradiamond_posts}
\end{figure*}

\begin{figure*}[!h]
    \centering
    \includegraphics[width=0.512\linewidth]{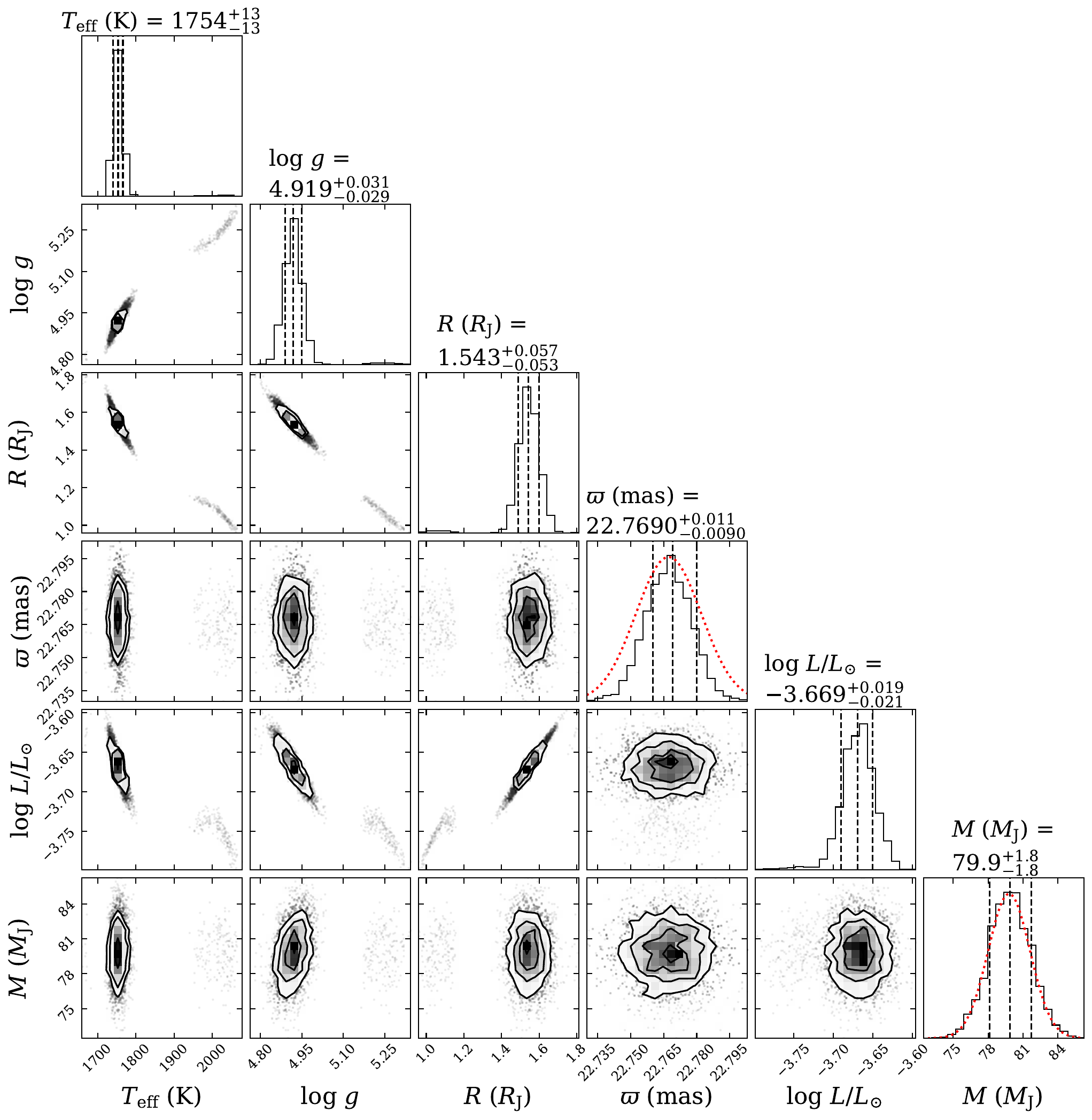}
    \caption{The posterior distributions of the physical and derived parameters from the BT-Settl atmospheric model fit to HD~206505~B (see Fig.~\ref{fig:hd206505_btsettl_main}).}
    \label{fig:hd206505_btsettl_posts}
\end{figure*}

\begin{figure*}[!h]
    \centering
    \includegraphics[width=0.61\linewidth]{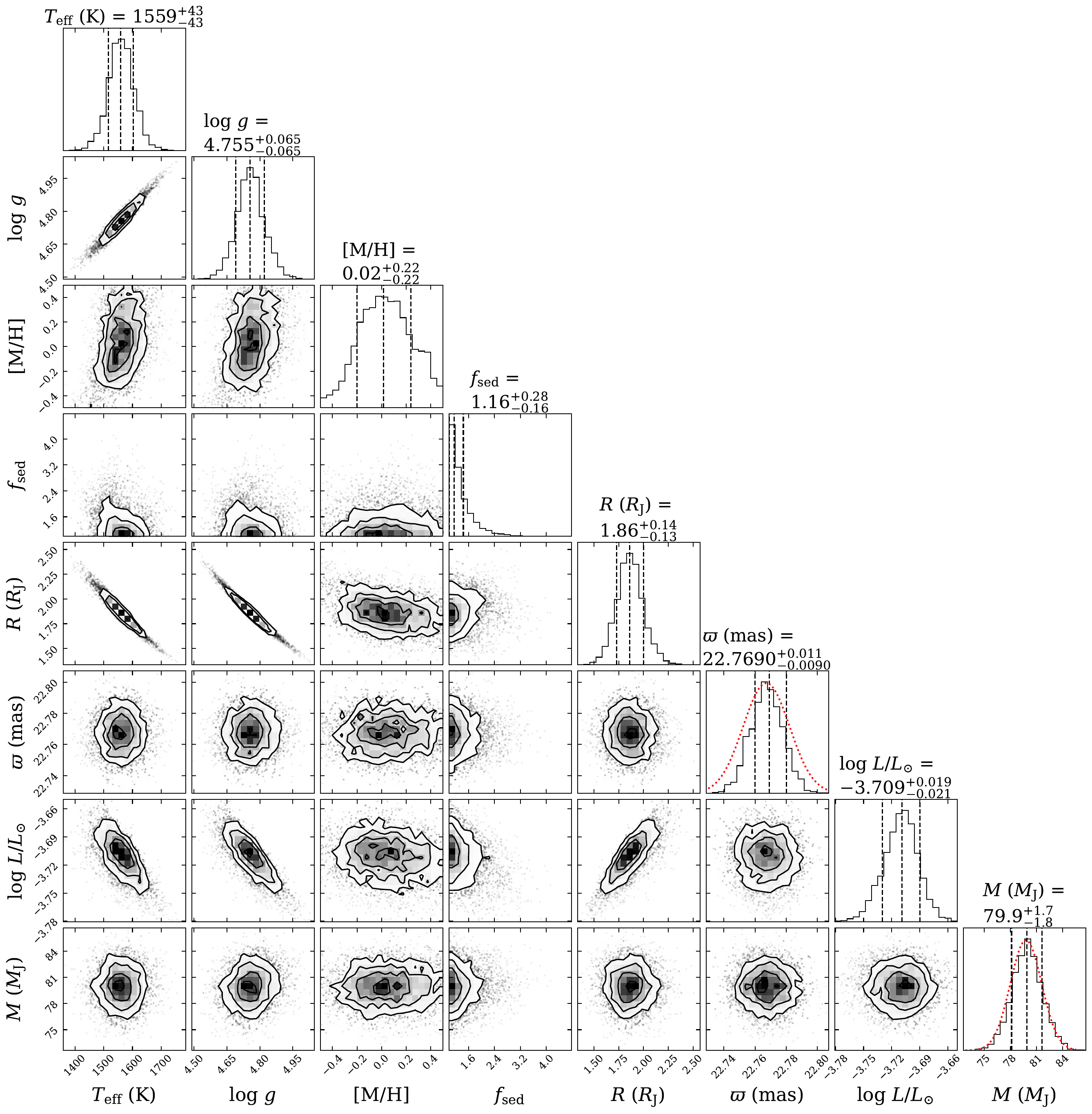}
    \caption{The posterior distributions of the physical and derived parameters from the Sonora Diamondback atmospheric model fit to HD~206505~B (see Fig.~\ref{fig:hd206505_sonoradiamond_main}).}
    \label{fig:hd206505_sonoradiamond_posts}
\end{figure*}

\end{appendix}

\end{document}